\documentclass{sig-alternate}
\usepackage{url}

\usepackage{graphicx}
\usepackage{algorithm}
\usepackage[noend]{algpseudocode}
\usepackage{subfigure}
\usepackage{amssymb, amsmath, graphicx, charter, latexsym}
\usepackage{enumerate}
\usepackage{ragged2e}
\usepackage{mathtools}
\usepackage{tabu}
\usepackage{epstopdf}
\usepackage{multirow}
\usepackage{verbatim}
\usepackage[xcdraw]{xcolor}
\usepackage{lipsum}
\usepackage{isomath}
\usepackage{mdframed}
\usepackage{siunitx}
\usepackage{bm}
\usepackage[cal=cm]{mathalfa}
\usepackage{float}

\DeclareMathAlphabet{\mathpzc}{OT1}{pzc}{m}{it}
\DeclarePairedDelimiterX{\bkt}[1]{(}{)}{\let\given\sgiven #1}
\DeclarePairedDelimiterX{\sbkt}[1]{[}{]}{\let\given\sgiven #1}

\newtheorem{definition}{Definition}

\newtheorem{remark}{Remark}
\DeclareMathOperator{\caL}{\mathcal{L}}
\DeclareMathOperator{\caI}{\mathcal{I}}
\DeclareMathOperator{\caD}{\mathcal{D}}
\DeclareMathOperator{\caV}{\mathcal{V}}

\DeclareMathOperator{\caG}{\mathcal{G}}
\DeclareMathOperator{\caM}{\mathcal{M}}
\DeclareMathOperator{\caU}{\mathcal{U}}

\DeclareMathOperator{\E}{\mathbb{E}}

\DeclareMathOperator{\T}{\intercal}
\newcommand{\bs}{\mathbf}
\newcommand\givenbase[1][]{\:#1\lvert\:}
\let\given\givenbase
\newcommand\sgiven{\givenbase[\delimsize]}
\newcommand{\norm}[1]{\left\lvert #1 \right \rvert}
\newcommand{\overbar}[1]{\mkern 1.5mu\overline{\mkern-1.5mu#1\mkern-1.5mu}\mkern 1.5mu}
\newmdtheoremenv[linewidth=1pt, skipabove=8pt, skipbelow=8pt]{theorem_md}{Theorem}
\newmdtheoremenv[linewidth=1pt, skipabove=8pt, skipbelow=8pt]{lemma_md}{Lemma}
\makeatletter
\newenvironment{breakablealgorithm}
  {
   \begin{center}
     \refstepcounter{algorithm}
     \hrule height.8pt depth0pt \kern2pt
     \renewcommand{\caption}[2][\relax]{
       {\raggedright\textbf{\ALG@name~\thealgorithm} ##2\par}%
       \ifx\relax##1\relax 
         \addcontentsline{loa}{algorithm}{\protect\numberline{\thealgorithm}##2}%
       \else 
         \addcontentsline{loa}{algorithm}{\protect\numberline{\thealgorithm}##1}%
       \fi
       \kern2pt\hrule\kern2pt
     }
  }{
     \kern2pt\hrule\relax
   \end{center}
  }
\makeatother

\begin{document}
\title{Throughput-Optimal Scheduling for Multi-Hop Networked Transportation Systems With Switch-Over Delay}
\author{Ping-Chun Hsieh, Xi Liu, Jian Jiao, I-Hong Hou, Yunlong Zhang, P. R. Kumar\\
       \affaddr{Texas A\&M University}\\
       \affaddr{\{pingchun.hsieh, xiliu, jiaojian, ihou, yz61, prk\}@tamu.edu}\\
} 
\maketitle
\begin{abstract}
The emerging connected-vehicle technology provides a new dimension in developing more intelligent traffic control algorithms for signalized intersections in networked transportation systems.
An important challenge for the scheduling problem in networked transportation systems is the switch-over delay caused by the guard time before any traffic signal change.
The switch-over delay can result in significant loss of system capacity and hence needs to be accommodated in the scheduling design. 
To tackle this challenge, we propose a distributed online scheduling policy that extends the well-known Max-Pressure policy to address switch-over delay by introducing a bias factor toward the current schedule.
We prove that the proposed policy is throughput-optimal with switch-over delay.
Furthermore, the proposed policy remains optimal when there are both connected signalized intersections and conventional fixed-time ones in the system.
With connected-vehicle technology, the proposed policy can be easily incorporated into the current transportation systems without additional infrastructure.
Through extensive simulation in VISSIM, we show that our policy indeed outperforms the existing popular policies.
\end{abstract} 

\section{Introduction}
\label{section:introduction}
Traffic congestion in urban area has been an increasingly severe problem in all cities of different sizes. According to a recent study~\cite{schrank20152015}, every driving commuter in the U.S. spends on average 30 to 60 hours of extra time on the road each year. Furthermore, about two thirds of the extra time comes from road congestion. 
For an urban transportation network which consists of intersections as nodes and roads between intersections as edges, intersections are often the source of road congestion as well as the accident-prone area~\cite{lomax1997quantifying}. 

Recently, considerable works are exploring novel scheduling strategies for intersections from the perspective of \emph{networked transportation systems}, which incorporate the emerging connected-vehicle technologies such as vehicle-to-vehicle (V2V) communication and vehicle-to-infrastructure (V2I) communication.
With connected-vehicle technologies, infrastructures can obtain accurate and real-time information about the number of vehicles waiting in each lane~\cite{tiaprasert2015queue}. 
The scheduling problem in networked transportation systems then becomes very similar to that in computer networks. 
In this analogy, each intersection corresponds to a router, each lane corresponds to a queue, and each vehicle corresponds to a packet. 
Indeed, there have been some efforts to apply the well-known Max-Pressure policy in computer networks \cite{tassiulas1992stability} to networked transportation systems \cite{varaiya2013max}.

Currently, most scheduling algorithms manage traffic flows at intersections via traffic signals,
whose color switches periodically between red and green. 
When the color is green, the traffic flow along the corresponding direction obtains the access to the intersection. 
The access will be revoked when the color switches to red. 
Transition from green phase to red phase is not instantaneous, but requires sufficient guard time for safety, which usually lasts for 3-8 seconds~\cite{lammer2008self}. 
The throughput during this transition phase is nearly zero. 
In addition, there is also throughput loss when a new green phase starts or ends because of acceleration or deceleration of vehicles. 
We capture such capacity loss by introducing \emph{switch-over delay} in this paper. 
The switch-over delay needs to be explicitly addressed in designing scheduling policies for intersections.
Unfortunately, most of the existing literature on scheduling of intersections via traffic signals ignore the effect of switch-over delay. 
In fact, Ghavami {\it et al.} ~\cite{ghavami2012delay} demonstrate that, while the dynamic signal control policies like the Max-Pressure policy outperform the conventional fixed-time policy in general, the performance of the dynamic signal control policies can be seriously affected when capacity loss due to switch-over delay is considered. 

Furthermore, during the transition between traditional transportation system and a fully connected system, only part of the intersections are equipped with sensors and V2I/V2V communication~\cite{liu2015towards}, while other intersections need to rely on conventional fixed-time control policies. 
In such partially-connected systems, the newly proposed policies are required to coexist well with conventional ones.

This paper aims to address all the above challenges. 
We propose a distributed scheduling policy for networked transportation systems and formally prove that the proposed policy is throughput-optimal under the existence of switch-over delay. 
The proposed policy accommodates the switch-over delay by adding a bias factor toward to the current schedule.
Moreover, we introduce a superframe structure which achieves synchronization among connected intersections and serves as a natural structure for stability analysis.
Our main contribution can be summarized as follows:
\begin{itemize}
\vspace{-2mm}
\item Switch-over delay is considered and tackled in the proposed policy, which is proved to be throughput-optimal under the existence of switch-over delay. 
\vspace{-2mm}
\item The proposed policy is distributed with low implementation complexity, and therefore it scales well with network size.
\vspace{-2mm}
\item Throughput-optimality of the proposed policy does not depend on any knowledge of traffic demands.
\vspace{-2mm}
\item Throughput-optimality of the proposed policy is preserved when there are both connected intersections and fixed-time intersections in the system. Therefore, the proposed policy can still perform well in partially-connected transportation systems.
\vspace{-2mm}
\item We evaluate the proposed policy via realistic microscopic simulation on a standard simulator for transportation research. 
\vspace{-2mm}
\end{itemize} 

While this paper focuses on networked transportation systems, our theoretical results are also applicable to many other applications with switch-over delay, such as optical networks \cite{Mcgarry2008}, wireless networks with directional antennas \cite{Navda2007}, and multi-thread operating systems \cite{David2007}.
The rest of the paper is organized as follows. 
Section \ref{section:model} describes the model of intersections and multi-hop transportation systems. 
The proposed scheduling policy is illustrated and the proof of optimality is provided in Section \ref{section:policy}. 
Section \ref{section:simulation} presents the simulation results.
Section \ref{section:conclusion} concludes the paper.

\section{Related Work}
\label{section:related}
In current transportation systems, traffic signals are often adaptively controlled by proprietary traffic control suites, such as SCATS \cite{Lowrie1982} and SCOOT \cite{SCOOT}.
Following the fixed-time control paradigm, these software suites require real-time traffic statistics to optimize cycle splits and offsets in the timing plan for some given objective functions.
However, since traffic demands can change rapidly with time, it might be difficult and costly to collect the statistics in a timely manner.

Different from the fixed-time approach, scheduling design based on real-time queue length information is attracting more and more attention due to the recent progress in connected-vehicle technology.
For example, adaptive control based on queue length is proposed in~\cite{tiaprasert2015queue}, where queue length is estimated via probe vehicles with V2I and V2V communication. 
On the other hand, inspired by the results in computer networks~\cite{tassiulas1992stability}, Varaiya \cite{varaiya2013max} and Wongpiromsarn {\it et al.}~\cite{wongpiromsarn2012distributed} propose individually a Max-Pressure policy for signal control and formally prove that the Max-Pressure policy is throughput-optimal when the queue capacity is infinite and the routing rates are known. 
To relax the assumption of infinite queue capacity, Xiao {\it et al.}~\cite{xiao2014pressure} present a variation of Max-Pressure policy that is throughput-optimal within a reduced capacity region when the queue capacity is finite but large enough.
To relax the assumption on routing rates, Gregoire {\it et al.}~\cite{gregoire2014back} also propose a back-pressure-based signal control policy and prove that it is throughput-optimal with unknown routing rates.
Despite the above progress, none of these policies takes the switch-over delay into account.

Among the existing literature on the scheduling design for systems with switch-over delay, \cite{Armony2003,Hung2008,Chan2016,Celik2016} are the most relevant to the scope of this paper.
First, Armony and Bambos \cite{Armony2003} study a system of parallel queues with switch-over delay and propose a family of dynamic cone policies and batch policies to achieve optimal throughput.
Subsequently, Hung and Chang \cite{Hung2008} present a generalized version of dynamic cone policy to reduce the complexity of the original cone policy. 
Chan \cite{Chan2016} also presents a Max-Weight type policy with hysteresis and prove that it is throughput-optimal for a system of parallel queues with deterministic service processes.
Besides, Celik {\it et al.} \cite{Celik2016} propose a family of generalized Max-Weight policies and prove that any policy satisfying the proposed criteria is throughput-optimal.
As an exemplar policy in \cite{Celik2016}, the Variable Frame-Based Max-Weight (VFMW) policy introduces a frame structure to avoid excessive capacity loss due to switch-over delay.
However, all of the above policies are designed specifically for single-hop systems and hence the optimality results may not be guaranteed in multi-hop systems.
In this paper, we regard VFMW as the reference policy for comparison in simulation.
In Section \ref{section:simulation}, we will show that the VFMW policy, which is throughput-optimal for single-hop systems, can actually perform poorly in multi-hop systems.

\section{System Model}
\label{section:model}
We model a multi-hop transportation system by a directed graph $(\caV,\caL)$, where $\caV$ denotes the set of intersections and $\caL$ is the set of directional links connecting the intersections.  
Each link has a start node and an end node.
In this paper, we use the terms \emph{node} and \emph{intersection} interchangeably.
For convenience, we also include one common virtual source node $v_{\textrm{s}}$ as well as one common virtual destination node $v_{\textrm{d}}$ in the directed graph.
We assume time is slotted. 
The links can be further divided into three categories: internal links $\caL_{\textrm{int}}$, entry links $\caL_{\textrm{entry}}$, and exit links $\caL_{\textrm{exit}}$. 
Each entry link has the same start node $v_{\textrm{s}}$ and an end node $v\in \caV$ where $v\neq v_{\textrm{d}}$. 
Similarly, each exit link has the same end node $v_{\textrm{d}}$ and a start node $v\in \caV$ where $v\neq v_{\textrm{s}}$.
Therefore, entry links and exit links together characterize the boundary of a system.
This model can also take garages into account by modelling each garage as an entry link plus an exit link.

\begin{figure}[H]
\begin{center}
\includegraphics[scale=0.29]{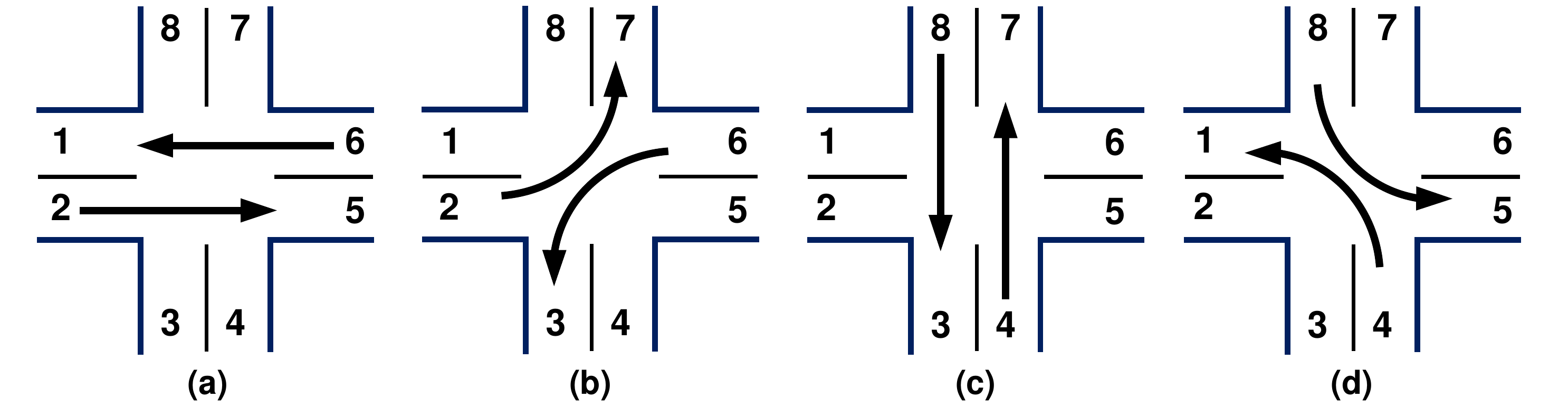}
\end{center}
\caption{A typical intersection with eight movements and 4 admissible phases.}
\label{figure:intersection}
\end{figure}

Given two links $i,j\in\caL$ incident to the same intersection, link $i$ is called a downstream link of $j$ (or equivalently, $i$ is an upstream link of $j$) if the end node of link $i$ is the same as the start node of link $j$. We use $\caD(i)$ and $\caU(i)$ to denote the set of all the downstream links and the set of all the upstream links of each link $i$, respectively.
Without loss of generality, we suppose that each link has at most $U_{\max}$ upstream links.
Moreover, the link pair $(i,j)$ forms a \emph{movement} of vehicles. 
We denote $\caM_v$ to be the set of movements of each intersection $v\in \caV$ and define $\caM:=\cup_{v\in \caV} \caM_v$. 
Besides, a collection of non-conflicting movements is called an \emph{admissible phase} of an intersection. 
For example, Figure \ref{figure:intersection} shows a standard intersection of eight movements and four admissible phases. 
In this typical intersection, each link has two upstream links and two downstream links.
For ease of explanation, we assume that vehicles can only go straight or turn left, but cannot turn right, in this example. 
Each movement $(i,j)$ has an associated queue $Q_{i,j}$ holding incoming vehicles. 
In other words, we assume that there exists a separate queue for each left-turn and through movement.
We assume that each queue has infinite size such that there is no overflow or blockage at each intersection.
Throughout this paper, we use the three-tuple $\caG=(\caV, \caL, \caM)$ to denote a transportation system.

External vehicles enter the system only via the entry links. 
For any entry link $i$ and its downstream link $j\in \caD(i)$, let $\{A_{i,j}(t)\}_{t\geq 0}$ be an i.i.d. sequence of external arrivals at $Q_{i,j}$ with average external arrival rate $\lambda_{i,j}>0$ and $A_{i,j}(t)\leq A_{\max}$ at any time $t$.
For any non-entry link $i$ and $j\in\caD(i)$, we simply let $A_{i,j}(t)=0$ for all $t$ and hence $\lambda_{i,j}=0$.
For ease of later discussion, we also define $\lambda_i:=\sum_{j\in\caD(i)}\lambda_{i,j}$ to be the total external arrival rate through each link $i$.
Similarly, let $\{S_{i,j}(t)\}_{t\geq 0}$ be an i.i.d. sequence of potential service rates of the movement $(i,j)$ with average service rate $\mu_{i,j}$, for each movement $(i,j)\in \caM$.
We also assume that $S_{i,j}(t)\leq S_{\max}$, for any movement $(i,j)$ and any time $t$. 
$S_{i,j}(t)$ captures the variation in the passage time required by different vehicles.
Since $S_{i,j}(t)$ depends on the instantaneous conditions such as vehicle speed and driver behavior, it is difficult for the traffic scheduler to obtain the information about potential service rates. 
Therefore, we presume that the traffic scheduler only has the information of \emph{average service rate}, which is often called \emph{saturation flow} in the transportation community.
Besides, the average service rate of a movement is roughly proportional to the number of lanes of that movement \cite{HCM2000}.

In our multi-hop model, vehicles are routed in a probabilistic manner.
When a vehicle enters a link $i$, it will choose to join a downstream link $j\in \caD(i)$ independently with probability $r_{i,j}$ with $\sum_{j\in \caD(i)} r_{i,j}=1$. 
Here we assume the routing probability $r_{i,j}>0$, for all movements $(i,j)\in \caM$.
Let $R_{i,j}(t)$ denote the portion of vehicles that join $Q_{i,j}$ among the vehicles entering link $i$ at time $t$, where $0\leq R_{i,j}(t)\leq 1$. 
Since each vehicle chooses its route independently, then we know that $\E[R_{i,j}(t)]=r_{i,j}$ for any time $t$ by the basic properties of multinomial random variables.
Note that the above model of arrivals, service, and routing is similar to that of the classic open Jackson network.

For each intersection, at each time slot exactly one of the admissible phases is chosen to have the right of way based on its scheduling policy. 
Let $I_{i,j}(t)$ be the indicator function of whether $Q_{i,j}$ is scheduled at the corresponding intersection at time $t$.
Therefore, for each intersection $v\in \caV$, we can use a $|\mathcal{M}_v|$-dimensional binary vector to represent the scheduled phase of the intersection. 
Let $\caI_v$ be the collection of the schedule vectors of all the admissible phases at the intersection $v$. 
Then, under a scheduling policy, each intersection $v$ determines $\bs{I}_v(t)\in \caI_v$ at each time $t$.

Moreover, in order to guarantee absolute safety, it takes non-zero time for an intersection to switch the right of way from the current schedule to the next. 
Such loss of service time during traffic signal change is modelled as \emph{switch-over delay}, during which all the movements at the intersection are prohibited and hence the throughput is zero.
For simplicity, we assume that the switch-over delay is $T_S$ slot(s) for all the intersections. 
Besides, an intersection is said to be \emph{active} if it is not in switch-over.
Let $X_{i,j}(t)$ be the indicator function of the event that the movement $(i,j)$ is active at time slot $t$. 
For each intersection, the time between two switch-over events is called a \emph{frame}.

In this paper, each intersection is either a {\it fixed-time intersection} or a {\it connected intersection}. 
For a fixed-time intersection, it simply follows the weighted round-robin policy with the weights determined a priori according to long-term average traffic demands.
In contrast, a connected intersection dynamically makes scheduling decisions based on real-time information obtained via connected-vehicle technology, such as queue length.
We use $\caV_{{F}}$ and $\caV_{C}$ to denote the set of fixed-time intersections and connected intersections, respectively.

For simplicity of notation, we use boldface fonts for vectors and matrices throughout the paper. For example, $\bm{\lambda}=(\lambda_{i})_{i\in \caL}$ denotes the per-link external arrival rate vector and ${\bf{Q}}(t)=(Q_{i,j}(t))_{(i,j)\in \caM}$ denotes the queue length vector of all the queues in the system.

\section{Capacity Region}
\label{section:capacity}
To study throughput-optimality, we first need to characterize the capacity region of a multi-hop transportation system.
\begin{definition}
A multi-hop transportation system is {strongly stable} under a scheduling policy $\pi$ if 
\begin{equation}
\limsup_{T\rightarrow\infty} \frac{1}{T} {\sum_{\tau=0}^{T-1} {\sum_{(i,j)\in \caM}\E\sbkt[\big]{Q_{i,j}(\tau)}}}< \infty.
\end{equation}
Meanwhile, we say that the policy $\pi$ {stabilizes} the system.
\end{definition}
Next, we introduce the definition of feasible external arrival rate vectors:
\begin{definition}
Given a multi-hop transportation system $\caG$, an external arrival rate vector $\bm{\lambda}=(\lambda_i)_{i\in\caL}$ is feasible if there exists a scheduling policy under which the system is strongly stable with $\bm{\lambda}$.
\end{definition}
From the definition of feasibility, we can define the capacity region as follows:
\begin{definition}
The capacity region is defined as the closure of the set of all the feasible external arrival rate vector $\bm{\lambda}$.
\end{definition}

To explicitly characterize the capacity region, we first obtain the {\it effective arrival rate}, which include both external arrivals and arrivals from upstream links, of each link and then provide the necessary and sufficient condition of the capacity region.
Let $\lambda_i^{*}$ be the effective arrival rate of link $i$.
According to our model, we have $\lambda_i^{*} = \lambda_i$ for all $i\in \caL_{\textrm{entry}}$. 
For any link $j\in \caL\setminus\caL_{\textrm{entry}}$, the effective arrival rate is determined by 
$\lambda_j^{*} = \sum_{i:j\in \caD(i)}{\lambda_i^{*}}r_{i,j}.$
Let $\bm{\lambda}^{*}=(\lambda_i^{*})_{i\in\caL}$ be the effective arrival rate vector and $\bs{R}=(r_{i,j})_{i,j\in\caL}$ be the routing probability matrix. Then, we can write the system of traffic equations in matrix form:
\begin{equation}
\bm{\lambda}^{*}=\bm{\lambda}+\bs{R}^{\T}\bm{\lambda}^{*},\label{equation:traffic equation matrix form}
\end{equation}
where $\bs{R}^{\T}$ is the transpose of the routing probability matrix.
Note that (\ref{equation:traffic equation matrix form}) is similar to the system of traffic equations of an open Jackson network.
Let $\bs{1}$ be an $\norm{\caL}\times \norm{\caL}$ identity matrix.
It is easy to verify that the equation in (\ref{equation:traffic equation matrix form}) has a unique solution as $\bm{\lambda}^{*}=(\bs{1}-\bs{R}^{\T})^{-1}\bm{\lambda}$, where $(\bs{1}-\bs{R}^{\T})$ is invertible (Section 2.1 in \cite{Chen2001}).

For each fixed-time intersection $v$, let $\xi_v\in (0,1)$ be the average fraction of time in which the intersection $v$ is in switch-over. 
Since the policy of each fixed-time intersection is given a priori, then $\xi_v$ is also fixed.
Let $\mathrm{\Lambda}$ be the set of all the external arrival rate vectors $\bm{\lambda} $ with which the following conditions hold:
(i) For each fixed-time intersection $v$, there exists $\epsilon>0$ and a vector ${\bm{\Sigma}_v}=(\Sigma_{i,j})_{(i,j)\in \caM_v}$ in the convex hull of $\caI_v$ such that the effective arrival rates satisfy that
\begin{equation}
\xi_{v}\mu_{i,j}\Sigma_{i,j}>\lambda_{i}^{*} r_{i,j}+\epsilon, \hspace{12pt} \forall (i,j)\in \caM_{v}.
\end{equation} 
In other words, a fixed-time intersection $v$ needs to have at least a small service margin for every movement at $v$.

(ii) For each connected intersection $v\in \caV_{C}$ there exists $\epsilon>0$ and a vector ${\bm{\Sigma_v}}=(\Sigma_{i,j})_{(i,j)\in \caM_v}$ in the convex hull of $\caI_v$ such that
\begin{equation}
\mu_{i,j}\Sigma_{i,j}>\lambda_{i}^{*} r_{i,j}+\epsilon, \hspace{12pt} \forall (i,j)\in \caM_{v}.
\end{equation}
Besides, let $\overbar{\mathrm{\Lambda}}$ be the closure of $\mathrm{\Lambda}$. 
The following Theorem \ref{theorem:feasible lambda} provides a sufficient condition for capacity region.
\begin{theorem_md}
\label{theorem:feasible lambda}
For a multi-hop transportation system with switch-over delay, an external arrival rate vector $\bm{\lambda}=(\lambda_i)_{i\in \caL}$ is feasible if $\bm{\lambda}\in \mathrm{\Lambda}$.
\end{theorem_md}
\begin{proof}
This can be proved by finding a proper fixed-time policy for each connected intersection. By Theorem 1 in \cite{varaiya2013max}, we directly know that given any $\bm{\lambda}\in \mathrm{\Lambda}$, there exists a fixed-time policy for each connected intersection such that the whole system is strongly stable. 
Hence, $\lambda$ must be feasible if $\lambda\in \mathrm{\Lambda}$. $\hfill$
\end{proof}
Next, we provide a necessary condition for capacity region in Theorem \ref{theorem:infeasible lambda}.
\begin{theorem_md}
\label{theorem:infeasible lambda}
For a multi-hop transportation system with switch-over delay and with an external arrival rate vector $\bm{\lambda}$, if $\bm{\lambda}\notin \overbar{\mathrm{\Lambda}}$, then there exists no policy under which the system is strongly stable.
\end{theorem_md}
\begin{proof}
This is a direct result of Theorem 1 in \cite{varaiya2013max}. $\hfill$
\end{proof}
Hence, by Theorem \ref{theorem:feasible lambda} and Theorem \ref{theorem:infeasible lambda}, the capacity region can be characterized as follows:

\begin{theorem_md}
Given a multi-hop transportation system $\caG$ with switch-over delay, the capacity region of $\caG$ is $\overbar{\mathrm{\Lambda}}$.
\end{theorem_md}
Given the knowledge of the capacity region, the concept of throughput-optimality is defined as follows:
\begin{definition}
Given a multi-hop transportation system $\caG$, a scheduling policy $\pi$ is said to be throughput-optimal if the system is strongly stable under $\pi$ with any external arrival rate vector $\bm{\lambda}\in \mathrm{\Lambda}$.
\end{definition}

\section{Scheduling For Throughput Optimality}
\label{section:policy}
In this section, we introduce our scheduling policy for connected intersections and prove that it is throughput-optimal with switch-over delay.

\subsection{A Throughput-Optimal Scheduling Policy}
\label{section:policy:policy}
To begin with, we define \emph{pressure} as follows:
\vspace{-1mm}
\begin{definition}
For any time $t$, the {pressure} of a movement $(i,j)\in \caM$ is defined as the difference between the queue length of $(i,j)$ and the weighted average of the queue lengths of $(j,k)$ for every $k\in \caD(j)$, i.e.
\begin{equation}
W_{i,j}(t):=Q_{i,j}(t)-\sum_{k:k\in \caD(j)} r_{j,k}Q_{j,k}(t). 
\end{equation}
In addition, for any intersection $v$, the pressure of any admissible phase $\bs{I}_{v}=(I_{i,j})\in\caI_v$ is defined as $\sum_{{i,j}\in\caM_v}\mu_{i,j}I_{i,j}W_{i,j}(t)$.
\end{definition}
We also introduce a useful definition:
\begin{definition}
A scheduling policy $\pi$ is said to be max-pressure-at-switch-over if $\pi$ always schedules the phase with the maximum pressure at each switch-over event. 
\end{definition}

Now, we formally present the \emph{Biased Max-Pressure (B-MP) scheduling policy} in Algorithm \ref{alg:B-MP}.
In B-MP, time is divided into consecutive {\it superframes}. At the beginning of a superframe, the duration of a superframe is calculated by (\ref{equation:BMP superframe size}). Whenever a connected intersection switches, it always switches to the phase with the maximum pressure, and therefore B-MP is max-pressure-at-switch-over. A connected intersection will only switch under two conditions: (i) at the beginning of each superframe, or (ii) when condition specified by (\ref{equation:B-MP LHS}) and (\ref{equation:B-MP RHS}) is satisfied. From condition  (\ref{equation:B-MP LHS})-(\ref{equation:B-MP RHS}), we can see that B-MP only make a switch when the maximum pressure is larger than the pressure of the current phase by a certain portion. Condition (\ref{equation:B-MP LHS}) can be interpreted as adding a bias factor toward the pressure of the current phase, and hence the name B-MP. This bias toward the current phase is to prevent the traffic signal from significant capacity loss due to frequent switch-overs.

Moreover, within one superframe, each connected intersection under B-MP can make scheduling decisions independently based on only the local queue length information. 
Therefore, B-MP is fully distributed within each superframe and the coordination among the connected intersections is minimal.
We use $t_{k}$ to denote the beginning of the $k$-th superframe and set $t_{0}=0$.
Let $T_{k}:=t_{k+1}-t_{k}$ be the length of the $k$-th superframe.
Besides, let $M^{v}_k$ be the number of switch-over events in the $k$-th superframe for each connected intersection $v$.
Since each superframe may contain different number of frames at different connected intersections, we use $t_{k,l}^{v}$ to denote the time of the $l$-th switch-over at intersection $v$ in the $k$-th superframe and set $t_{k,0}^{v}=t_k$.

\begin{breakablealgorithm}
\caption{Biased Max-Pressure Policy (B-MP)}\label{alg:B-MP}
\begin{algorithmic}[1]
\State At time $t=t_k$, obtain the length of the $k$-th superframe:
\begin{equation}
\label{equation:BMP superframe size}
T_k=\bkt[\Big]{\sum_{(i,j)\in \caM}Q_{i,j}(t_k)}^{\beta}, \hspace{6pt}\beta\in (0,1). 
\end{equation}
Calculate the beginning of the next superframe as $t_{k+1}=t_{k}+T_k$.
\State Find the phase with the largest pressure at current time $t$, i.e.
\begin{equation*}
\bs{I}_{v}^{*}(t)\in{\arg\max}_{\bs{I}\in\caI_v} \sum_{(i,j)\in\caM_v}\mu_{i,j}I_{i,j}W_{i,j}(t).
\end{equation*}
Ties are broken arbitrarily.
\vspace{1mm}
\State If $\bs{I}_{v}^{*}(t)\neq \bs{I}_{v}^{*}(t-1)$, initiate switch-over for the next $T_S$ slots and then apply the new schedule $\bs{I}_{v}^{*}(t)$ for one slot. Else, directly apply $\bs{I}_{v}^{*}(t)$ for one slot. 
\vspace{1mm}
\State For any $t\in [t_{k,l}^{v}, t_{k,l+1}^{v})$ in the rest of the $k$-th superframe, find the phase $\bs{I}_{v}^{*}(t)$ that has the largest pressure. If the intersection is not in switch-over at time $t$, the intersection will make a switch if the following condition is satisfied:
\begin{align}
\bkt[\Big]{1+B_{v}(t_{k,l}^{v})}\biggl(&\sum_{(i,j)\in\caM_v}\mu_{i,j}I^{*}_{i,j}(t-1)W_{i,j}(t)\biggr)^{+}\label{equation:B-MP LHS}\\
&<\bkt[\bigg]{\sum_{(i,j)\in\caM_v}\mu_{i,j}{I}^{*}_{i,j}(t)W_{i,j}(t)}^{+},\label{equation:B-MP RHS}
\end{align}  
where $x^{+}$ is a shorthand for $\max\{x,0\}$ and $B_{v}(\cdot)$ is the bias function defined as
\begin{equation}
\label{equation:bias function definition}
B_{v}(t)={\zeta T_S}{\min\Biggl\{1, \biggl(\biggl[\sum_{(i,j)\in\caM_v}W_{i,j}(t)\biggr]^{+}\biggr)^{-\alpha}\Biggr\}}
\end{equation}
with $\alpha\in (0,1)$ and $\zeta>0$.
Otherwise, stay at the current phase.
\vspace{1mm}
\State Repeat Step 3 and Step 4 until the end of the $k$-th superframe.
\vspace{1mm}
\State At $t=t_{k+1}$, go back to step 1 and repeat the above procedure for the next superframe.
\end{algorithmic}
\end{breakablealgorithm}
\vspace{2mm}

\subsection{Proof of Throughput-Optimality}
\label{section:policy:proof}

To study system stability, we consider the queue length update over one superframe. Define $\mathrm{\Delta} Q_{i,j}(t_k):=Q_{i,j}(t_{k+1})-Q_{i,j}(t_{k})$. 
For any movement $(i,j)$ with link $i\in \caL_{\text{entry}}$, we have
\begin{align}
&\mathrm{\Delta} Q_{i,j}(t_k)\\
&=-\sum_{t=t_{k}}^{t_{k+1}-1}\bkt[\Big]{S_{i,j}(t)I_{i,j}(t)X_{i,j}(t)\wedge Q_{i,j}(t)}+\sum_{t=t_{k}}^{t_{k+1}-1}A_{i,j}(t),\label{equation:entry queue evolution}
\end{align}
where $(x\wedge y)$ is a shorthand for $\min\{x,y\}$. 
Note that the first term of (\ref{equation:entry queue evolution}) represents the number of vehicles that actually leaves $Q_{i,j}$ during the $k$-th superframe and the second term is the total external arrivals at $Q_{i,j}$ in the $k$-th superframe.

On the other hand, for any movement $(i,j)\in \caM$ with link $i\notin \caL_{\textrm{entry}}$, we have
\begin{align}
&\mathrm{\Delta} Q_{i,j}(t_k)=-\sum_{t=t_{k}}^{t_{k+1}-1}\bkt[\Big]{S_{i,j}(t)I_{i,j}(t)X_{i,j}(t)\wedge Q_{i,j}(t)}\label{equation:nonentry queue evolution 1}\\
&+\sum_{t=t_{k}}^{t_{k+1}-1}\sum_{m:(m,i)\in \caM}\bkt[\Big]{S_{m,i}(t)I_{m,i}(t)X_{m,i}(t)\wedge Q_{m,i}(t)}R_{i,j}(t).\label{equation:nonentry queue evolution 2}
\end{align} 

Note that (\ref{equation:nonentry queue evolution 2}) represents the total number of vehicles coming from the upstream links of $i$ during the $k$-th superframe.

To study the throughput performance of such policies, we apply Lyapunov drift analysis and study the Lyapunov drift across one superframe. 
Define a Lyapunov function as
\begin{equation}
L({\bs{Q}}(t)):={\bs{Q}}(t)^{\T}{\bs{Q}}(t)=\sum_{(i,j)\in \caM}Q_{i,j}(t)^{2},
\end{equation}
where $\bs{Q}(t)^{\T}$ is the transpose of the queue length vector.
Define the Lyapunov drift over the $k$-th superframe as $\mathrm{\Delta} L(t_k):=L(\bs{Q}(t_{k+1}))-L(\bs{Q}(t_{k}))$. Then, we have
\begin{equation}
\mathrm{\Delta} L(t_k)=2{\bs{Q}}(t_k)^{\T}{\mathrm{\Delta}\bs{Q}(t_k)} + {\mathrm{ \Delta}\bs{Q}}^{\T}{\mathrm{\Delta}\bs{Q}(t_k)},
\end{equation} 
where $\mathrm{{\Delta}}\bs{Q}(t_k):=\bs{Q}(t_{k+1})-\bs{Q}(t_{k})$.
Given $\bs{Q}(t_{k})$, the size of the $k$-th superframe is known and therefore the conditional drift over the $k$-th superframe is well-defined. 
Note that it is actually not straightforward to calculate the conditional drift over one superframe:
\vspace{-1mm}
\begin{itemize}
\item For any intersection, there could be multiple frames and hence multiple phases scheduled in a stochastic sequence in one superframe. 
\vspace{-2mm}
\item Different intersections could possibly have totally different frame sizes in the same superframe.
\vspace{-2mm}
\item Given the queue length information at the beginning of a superframe, it is still not clear when switch-over will be triggered and which phase will be scheduled at each intersection since the arrival and service processes are stochastic.
\end{itemize}
\vspace{-2mm}
Despite the above challenges, the conditional drift over one superframe can still be characterized for the max-pressure-at-switch-over policies.
We first provide an upper bound on the conditional drift in the following lemma.

\begin{lemma_md}
\label{lemma:drift upper bound}
Given any $\bm{\lambda}\in \mathrm{\Lambda}$, under any max-pressure-at-switch-over policy with superframe structure, the conditional drift over one superframe is upper bounded as
\begin{align}
\E&\sbkt[\big]{\mathrm{\Delta} L(t_k)\sgiven \bs{Q}(t_k)}\leq -2\epsilon T_k \sum_{(i,j)\in\caM}{{W_{i,j}(t_k)}^{+}}\label{equation:upper bound term 1}\\
&+C_1 \sum_{v\in\caV_{C}} M_{k}^{v}\bkt[\bigg]{\sum_{(i,j)\in\caM_{v}}{W_{i,j}(t_k)}^{+}}\label{equation:upper bound term 2}\\
&+C_2\sum_{v\in \caV_{{F}}}\sum_{(i,j)\in \caM_{v}}{W_{i,j}(t_k)^{+}}+C_3 T_k^{2}+C_4 T_k\label{equation:upper bound term 3}
\end{align}
where $C_1$, $C_2$, $C_3$ and $C_4$ are finite positive constants and $x^{+}$ is a shorthand for $\max\{x,0\}$. 
\end{lemma_md}
\begin{proof}
With the max-pressure-at-switch-over property, we are able to quantify the pressure of the scheduled phases at any $t\in [t_k, t_{k+1})$ even if the scheduling decision of each frame is not known. The complete proof is provided in the Appendix \ref{appendix: proof of lemma 1}. $\hfill$
\end{proof}
\begin{remark}
Note that (\ref{equation:upper bound term 1}) represents the negative drift required for system stability. Besides, (\ref{equation:upper bound term 2}) and the first term of (\ref{equation:upper bound term 3}) represent the loss of service due to switch-over at connected intersections and the fixed-time intersections, respectively. The second and the third term of (\ref{equation:upper bound term 3}) stand for the service loss due to possible emptiness of the scheduled queues.
\end{remark}

\begin{remark}
Note that in (\ref{equation:upper bound term 2}) the service loss due to switch-over is basically a direct sum of the service loss contributed by each connected intersection. In other words, the performance of connected intersections are completely decoupled. Due to this feature, Lemma \ref{lemma:drift upper bound} still holds if different connected intersections follow different max-pressure-at-switch-over policies with superframe structure.
\end{remark}

To show that B-MP is throughput-optimal, we introduce a sufficient condition for strong stability in the following lemma.
\begin{lemma_md}
\label{lemma:condition for strong stability}
For any max-pressure-at-switch-over scheduling policy with superframe determined by (\ref{equation:BMP superframe size}), if there exists some constant $B_0>0$, $\epsilon_0>0$ and the conditional drift satisfies that
\begin{equation}
\E\sbkt[\big]{\mathrm{\Delta} L(t_{k})\sgiven \bs{Q}(t_k)}\leq B_0 -\epsilon_{0}\bkt[\bigg]{\sum_{(i,j)\in\caM}Q_{i,j}(t_{k})}^{1+\beta},\label{equation:drift upper bound in lemma}
\end{equation}
then we have
\begin{equation}
\limsup_{T\rightarrow \infty}\frac{1}{T}{\sum_{t=0}^{T-1}\sum_{(i,j)\in\caM}\E\sbkt[\big]{Q_{i,j}(t)}}< \infty.
\end{equation}
\end{lemma_md}
\begin{proof}
Define $H(t_k):=\sum_{t=0}^{T_k-1}\sum_{(i,j)\in \caM}Q_{i,j}(t_k+t)$. Then, we have
\begin{align*}
H(t_k)\leq& \sum_{t=0}^{T_k-1}\sum_{i\in \caL_{\textrm{entry}},j\in \caD(i)} \bkt[\Big]{Q_{i,j}(t_k)+\sum_{s=0}^{T_k-1}A_{i,j}(t_k+s)} \\
&+\sum_{t=0}^{T_k-1}\sum_{i\in \caL_{\textrm{int}},j\in \caD(i)} Q_{i,j}(t_k).
\end{align*}
After taking conditional expectation of $H(t_k)$, we have
\begin{align}
&\E\sbkt[\big]{H(t_k)\sgiven \bs{Q}(t_k)}\\
&\leq T_k^{2} \bkt[\bigg]{\sum_{i\in\caL_{\text{entry}}, j\in \caD(i)} \lambda_i^{*}r_{i,j}} + T_k \bkt[\bigg]{\sum_{(i,j)\in \caM}Q_{i,j}(t_k)}\\
&\leq B_1 \bkt[\bigg]{\sum_{(i,j)\in\ \caM}Q_{i,j}(t_k)}^{1+\beta}
\end{align}
where $B_1=1+\sum_{i\in\caL_{\text{entry}}}\lambda_i^{*}r_{i,j}$ is a positive constant independent of $\bs{Q}(t_k)$. Then, by (\ref{equation:drift upper bound in lemma}),
\begin{align}
\E\sbkt[\big]{\mathrm{\Delta} L(t_{k})\sgiven \bs{Q}(t_k)}\leq B_0-\frac{\epsilon_0}{B_1}\E\sbkt[\big]{H(t_k)\sgiven \bs{Q}(t_k)}.\label{equation:drift upper bound by H}
\end{align}
By summing (\ref{equation:drift upper bound by H}) over all the superframes, we have
\begin{align}
\sum_{k\geq 0}\E\sbkt[\big]{\mathrm{\Delta} L(t_{k})\sgiven \bs{Q}(t_k)}\leq \sum_{k\geq 0} \bkt[\Big]{B_0-\frac{\epsilon_0}{B_1}\E\sbkt[\big]{H(t_k)\sgiven \bs{Q}(t_k)}}.\label{equation:drift upper bound by H all}
\end{align}
Given a finite initial condition $\bs{Q}(0)$, we have $L(0)<\infty$ and $\sum_{k\geq 0}\E\sbkt[\big]{\mathrm{\Delta} L(t_{k})\sgiven \bs{Q}(t_k)}\geq -L(0)$. Hence, we conclude that
\begin{align*}
\limsup_{T\rightarrow \infty}\frac{\sum_{t=0}^{T-1}\E\sbkt[\big]{\sum_{(i,j)\in\caM}Q_{i,j}(t)}}{T}\leq \frac{B_1\bkt[\big]{B_0+L(0)}}{\epsilon_0}< \infty.
\end{align*}
$\hfill$
\end{proof}

Next, since Lemma \ref{lemma:drift upper bound} involves both queue length and pressure, we provide a useful inequality between total queue length and total pressure as follows.
\begin{lemma_md}
\label{lemma:sum W and sum Q inequality}
For any queue length vector $\bs{Q}=(Q_{i,j})$ and its corresponding pressure vector $\bs{W}=(W_{i,j})$, there must exist a constant $\delta>0$ such that 
\begin{equation}
\label{equation:sum W and sum Q inequality}
\sum_{(i,j)\in \caM}{W_{i,j}}^{+}\geq \delta\bkt[\bigg]{\sum_{(i,j)\in \caM}{Q_{i,j}}}.
\end{equation}
\end{lemma_md}
\begin{proof}
We provide a sketch of the proof: We first construct a new system by adding several dummy links and dummy movements to the original system and show that the new system is strongly connected and the corresponding routing matrix is invertible. By applying the Perron-Frobenius Theorem to the routing matrix, we obtain a strictly positive eigenvector with a positive eigenvalue.
Based on the eigenvector properties, we show that there must exist a constant $\delta>0$ such that the inequality (\ref{equation:sum W and sum Q inequality}) holds. The complete proof is provided in Appendix \ref{appendix: proof of lemma 3}.$\hfill$
\end{proof}

Note that B-MP is a max-pressure-at-switch-over policy and therefore Lemma \ref{lemma:drift upper bound} holds under the B-MP policy. 
To characterize the number of switch-over events in one superframe under the B-MP policy, we provide an upper bound on the size of each frame as follows.



\begin{lemma_md}
\label{lemma:Tkl lower bound under B-MP}
Under the B-MP policy, there exists a constant $C_5>0$ such that the length of each frame is lower bounded as
\begin{equation}
\label{equation:Tkl lower bound under B-MP}
T_{k,l}^{v}\geq {C_{5}}{B_{v}(t_{k,l}^{v})}\bkt[\Big]{\sum_{(i,j)\in \caM_v}{W_{i,j}(t_{k,l}^{v})^{+}}}.
\end{equation}
\end{lemma_md}

\begin{proof}
The proof is provided in Appendix \ref{appendix: proof of lemma 4}.$\hfill$
\end{proof}

With Lemma \ref{lemma:Tkl lower bound under B-MP}, we are ready to provide a lower bound on the number of switch-over events in one superframe under B-MP.

\begin{lemma_md}
\label{lemma:upper bound of Mk times sum W under B-MP}
For any intersection $v$ under the B-MP policy with bias function defined by (\ref{equation:bias function definition}),
we have $\forall k\geq 0$,
\begin{equation}
M_{k}^{v}\bkt[\Big]{\sum_{(i,j)\in \caM_{v}}{W_{i,j}(t_k)}^{+}}=\text{o}\bkt[\bigg]{\bkt[\Big]{\sum_{(i,j)\in \caM}Q_{i,j}(t_k)}^{1+\beta}}.
\end{equation}
\end{lemma_md}

\begin{proof}
The proof is provided in Appendix \ref{appendix: proof of lemma 5}. $\hfill$
\end{proof}

We are ready to show that B-MP is throughput-optimal.
\begin{theorem_md}
The B-MP policy is throughput-optimal for any $\alpha\in(0,1)$, $\beta\in (0,1)$.
\label{theorem:B-MP throughput-optimal}
\end{theorem_md}
\begin{proof}
Since B-MP is a max-pressure-at-switch-over policy with superframe structure, then Lemma \ref{lemma:drift upper bound} holds under B-MP. Therefore, by Lemma \ref{lemma:sum W and sum Q inequality} and the fact that $W_{i,j}(t)^{+}\leq Q_{i,j}(t)$ for any movement $(i,j)$ and any time $t$, we have
\begin{align}
\E&\sbkt[\big]{\mathrm{\Delta} L(t_k)\sgiven \bs{Q}(t_k)}\leq -2\epsilon \delta_{0}T_k \sum_{(i,j)\in\caM}{{Q_{i,j}(t_k)}}\label{equation:drift upper bound B-MP 1}\\
&+C_1 \sum_{v\in\caV_{C}} M_{k}^{v}\bkt[\Big]{ \sum_{(i,j)\in\caM_{v}}{W_{i,j}(t_k)}^{+}}\label{equation:drift upper bound B-MP 2}\\
&+C_2\sum_{v\in \caV_{{F}}}\sum_{(i,j)\in \caM_{v}}{Q_{i,j}(t_k)}+C_3 T_k^{2}+C_4T_k.\label{equation: drift upper bound B-MP 3}
\end{align}
By Lemma \ref{lemma:upper bound of Mk times sum W under B-MP} and the choice of $T_k$, we know $T_k \sum_{(i,j)\in\caM}{{Q_{i,j}(t_k)}}$ is the dominating term among (\ref{equation:drift upper bound B-MP 1})-(\ref{equation: drift upper bound B-MP 3}). Therefore, there exists a constant $B>0$ such that
\begin{align}
\E\sbkt[\big]{\mathrm{\Delta} L(t_k)\sgiven \bs{Q}(t_k)}\leq B - \epsilon\delta_{0}\bkt[\bigg]{\sum_{(i,j)\in\caM}{Q_{i,j}(t_k)}}^{1+\beta}.
\end{align}
By Lemma \ref{lemma:condition for strong stability}, we know that the system is strongly stable under the B-MP policy for any external arrival rate $\bm{\lambda}\in\mathrm{\Lambda}$. Hence, the B-MP policy is throughput-optimal. $\hfill$
\end{proof}

\begin{remark}
By Theorem \ref{theorem:B-MP throughput-optimal}, B-MP can achieve throughput-optimality with any $\alpha$ between 0 and 1. 
Meanwhile, the choice of $\alpha$ can indeed affect the average delay performance.
The issue on choosing $\alpha$ for achieving optimal delay in multi-hop systems with switch-over delay will be our future work and is beyond the scope of this paper.  
\end{remark}
\begin{remark}
The parameter $\beta$ determines the superframe size for coordination amoing the intersections. 
To minimize the coordination overhead, $\beta$ is recommended to be close to 1.
\end{remark}

\section{Extensions of Biased Max-Pressure Policy}
\label{section:variations}

\subsection{Weighted Queue Length}
\label{section:variations:weighted}
The concept of pressure can be further generalized by using \emph{weighted queue legnth}:
\begin{definition}
\label{definition:generalized pressure}
Let $q_{i,j}>0$ be the pre-determined weight factor of movement $(i,j)$. For each movement $(i,j)$, we define the weighted queue length as $\hat{Q}_{i,j}(t):=q_{i,j}Q_{i,j}(t)$, for all $t$. Then, the generalized pressure is defined as
\begin{equation}
\hat{W}_{i,j}(t):=\hat{Q}_{i,j}(t)-\sum_{k:k\in \caD(j)} r_{j,k}\hat{Q}_{j,k}(t). 
\end{equation}
\end{definition}

By substituting $\hat{W}_{i,j}(t)$ for $W_{i,j}(t)$, the B-MP policy remains throughput-optimal:
\begin{theorem_md}
The B-MP policy using the generalized pressure in Definition \ref{definition:generalized pressure} is still throughput-optimal for any $\alpha\in(0,1)$, any $\beta\in(0,1)$.
\end{theorem_md}
\begin{proof}
This can be proved by considering the drift of a Lyapunov function: $\hat{L}(\bs{Q}(t))=\sum_{(i,j)\in\caM}q_{i,j}Q_{i,j}(t)^{2}$. The rest of the proof is similar to that of Theorem \ref{theorem:B-MP throughput-optimal} and hence omitted due to space limitation. $\hfill$
\end{proof}

One important application of weighted queue length is to design a capacity-aware version of the B-MP policy in order to mitigate the \emph{queue overflow effect} due to finite queue capacity. Queue overflow often occurs when the system operates under oversaturated traffic (even merely for a short period of time). The overflow effect can lead to significant service loss as well as severe delay. Given the information about queue capacity, we can properly choose $q_{i,j}$ for each movement $(i,j)$ to reduce the chance of queue overflow. For example, choosing $q_{i,j}$ to be inversely proportional to the queue capacity of $Q_{i,j}$ is suggested in \cite{gregoire2015capacity}. In Section \ref{section:simulation}, we show an example of applying weighted queue length in simulation.  

\subsection{Estimated Queue Length With Bounded Error}
\label{section:variations:bounded error}
In networked transportation systems, it might be difficult or expensive to obtain completely accurate queue length information due to the latency in communication or random error in sensor detection. Let $Q_{i,j}^{\dagger}(t)$ and $W_{i,j}^{\dagger}(t)$ be the estimated queue length and the corresponding pressure, respectively. If the estimation error of queue length is always upper bounded, then the B-MP is still throughput-optimal with the estimated queue length.
We still consider the Lyapunov function $L({\bs{Q}}(t))=\sum_{(i,j)\in \caM}Q_{i,j}(t)^{2}$ and the corresponding drift conditioned on $\bs{Q}_{i,j}^{\dagger}(t_k)$. 
Then, we have the following upper bound on the conditional drift:
\begin{lemma_md}
\label{lemma:drift upper bound estimated}
Given any $\bm{\lambda}\in \mathrm{\Lambda}$, under the B-MP policy using estimated queue length $(Q_{i,j}^{\dagger}(t))$, if there exists a constant $B>0$ such that $\norm{Q_{i,j}(t)-Q_{i,j}^{\dagger}(t)}\leq B$ for all $(i,j)$ and all $t$, the conditional drift over one superframe is upper bounded as:
\begin{align}
\E&\sbkt[\big]{\mathrm{\Delta} L(t_k)\sgiven \bs{Q}^{\dagger}(t_k)}\leq -2\epsilon T_k \sum_{(i,j)\in\caM}{{W_{i,j}^{\dagger}(t_k)}^{+}}\label{equation:upper bound estimated term 1}\\
&+C^{\dagger}_1 \sum_{v\in\caV_{C}} M_{k}^{v}\bkt[\bigg]{\sum_{(i,j)\in\caM_{v}}{W_{i,j}^{\dagger}(t_k)}^{+}}\label{equation:upper bound estimated term 2}\\
&+C^{\dagger}_2\sum_{v\in \caV_{{F}}}\sum_{(i,j)\in \caM_{v}}{W^{\dagger}_{i,j}(t_k)^{+}}+C^{\dagger}_3 T_k^{2}+C^{\dagger}_4 T_k\label{equation:upper bound estimated term 3}
\end{align}
where $C^{\dagger}_1$, $C^{\dagger}_2$, $C^{\dagger}_3$ and $C^{\dagger}_4$ are finite positive constants. 
\end{lemma_md}
\begin{proof}
The proof is similar to that of Lemma \ref{lemma:drift upper bound}, and the main differences are: (i) Since the drift is now conditioned on $\bs{Q}^{\dagger}(t_k)$ instead of $\bs{Q}(t_k)$, the estimation error introduces an extra term in $\E\Bigl[{{\bs{Q}}(t_k)^{\T}{\mathrm{\Delta}\bs{Q}(t_k)}\given{\bs{Q}}^{\dagger}(t_k)}\Bigr]$. Due to the boundedness of estimation error, this extra term is at most of the same order as $T_k$. (ii) For connected intersections, B-MP using $\bs{Q}^{\dagger}(t_k)$ makes scheduling decisions based on $\bs{W}^{\dagger}(t_k)$. Therefore, B-MP is max-pressure-at-switch-over in terms of $\bs{W}^{\dagger}(t_k)$ instead of $\bs{W}(t_k)$. Besides, since $Q_{i,j}(t)-Q_{i,j}^{\dagger}(t)\in[-B, B]$, we also have $W_{i,j}(t)-W_{i,j}^{\dagger}(t)\in [-2B,2B]$, for all $(i,j)$ and all $t$. As a result, the bounded error in pressure only affects the coefficients of the existing terms in the original drift expression.
The complete proof is in Appendix \ref{appendix: proof of Lemma 6}. $\hfill$
\end{proof}
Now, we are ready to prove that B-MP is throughput-optimal with estimated queue lengths.
\begin{theorem_md}
If there exists a constant $B>0$ such that $\norm{Q_{i,j}(t)-Q_{i,j}^{\dagger}(t)}\leq B$ for all $(i,j)$ and all $t$, then B-MP is still throughput-optimal using the estimated queue length $(Q_{i,j}^{\dagger}(t))$.
\end{theorem_md}
\begin{proof}
First, we have $Q_{i,j}(t)-Q_{i,j}^{\dagger}(t)\in[-B, B]$ and $W_{i,j}(t)-W_{i,j}^{\dagger}(t)\in [-2B,2B]$, for all $(i,j)$ and all $t$, Besides, Lemma \ref{lemma:sum W and sum Q inequality} holds regardless of the scheduling policy. Therefore, we can rewrite the upper bound in Lemma \ref{lemma:drift upper bound estimated} as
\begin{align}
\E&\sbkt[\big]{\mathrm{\Delta} L(t_k)\sgiven \bs{Q}^{\dagger}(t_k)}\leq -2\epsilon \delta_{0}T_k \sum_{(i,j)\in\caM}{{Q_{i,j}^{\dagger}(t_k)}}\label{equation:drift upper bound B-MP estimated 1}\\
&+C^{\ddagger}_1 \sum_{v\in\caV_{C}} M_{k}^{v}\bkt[\Big]{ \sum_{(i,j)\in\caM_{v}}{W^{\dagger}_{i,j}(t_k)}^{+}}\label{equation:drift upper bound estimated B-MP 2}\\
&+C^{\ddagger}_2\sum_{v\in \caV_{{F}}}\sum_{(i,j)\in \caM_{v}}{W^{\dagger}_{i,j}(t_k)}+C^{\ddagger}_3 T_k^{2}+C^{\ddagger}_4T_k, \label{equation:drift upper bound estimated B-MP 3}
\end{align}
where $C^{\ddagger}_1$, $C^{\ddagger}_2$, $C^{\ddagger}_3$, $C^{\ddagger}_4$ are finite positive constants.
Furthermore, with a slight modification of the proof we know that Lemma \ref{lemma:Tkl lower bound under B-MP} and Lemma \ref{lemma:upper bound of Mk times sum W under B-MP} still hold when $\bs{W}(t_k)$ is replaced by $\bs{W}^{\dagger}(t_k)$ under B-MP.
By the same argument as that in the proof of Theorem \ref{theorem:B-MP throughput-optimal}, we know that $-2\epsilon T_k \sum_{(i,j)\in\caM}{{Q^{\dagger}_{i,j}(t_k)}}$ is the dominating term in (\ref{equation:drift upper bound B-MP estimated 1})-(\ref{equation:drift upper bound estimated B-MP 3}). 
Therefore, there must exist a constant $B^{\dagger}>0$ such that
\begin{align}
\E\sbkt[\big]{\mathrm{\Delta} L(t_k)\sgiven \bs{Q}^{\dagger}(t_k)}\leq B^{\dagger} - \epsilon\delta_{0}\bkt[\bigg]{\sum_{(i,j)\in\caM}{Q^{\dagger}_{i,j}(t_k)}}^{1+\beta}.\label{equation:drift upper bound in lemma estimated}
\end{align}
By the similar procedure as in Lemma \ref{lemma:condition for strong stability}, we know that (\ref{equation:drift upper bound in lemma estimated}) is also a sufficient condition for strong stability.
Hence, we conclude that B-MP remains throughput-optimal when the error in queue length is bounded. $\hfill$
\end{proof}

From Theorem \ref{theorem:B-MP throughput-optimal}, we know that B-MP is also robust to estimation error in queue length information.

\subsection{Limitations on Green Period}
\label{section:variations:green}
Conventionally, the timing plan of traffic signals includes a minimum green time to accommodate the vehicle startup delay. Under the B-MP policy, the minimum green time can be easily incorporated by introducing a minimum frame size $T_{G,\min}> T_S$. Accordingly, (\ref{equation:Tkl lower bound under B-MP}) in Lemma \ref{lemma:Tkl lower bound under B-MP} would be
\begin{equation}
T_{k,l}^{v}\geq \max\Biggl\{T_{G,\min},{C_{5}}{B_{v}(t_{k,l}^{v})}\bkt[\Big]{\sum_{(i,j)\in \caM_v}{W_{i,j}(t_{k,l}^{v})^{+}}}\Biggr\}.
\end{equation}
With a slight modification of the proof of Lemma \ref{lemma:upper bound of Mk times sum W under B-MP}, the B-MP policy with a minimum frame size still remains throughput-optimal.
On the other hand, a maximum green time is sometimes applied in the actuated version of fixed-time policy to avoid excessive delay of the minor roads. While this can also be included in B-MP by introducing a maximum frame size $T_{G,\max}$, setting a maximum frame size can result in loss of system throughput since the fraction of time spent on switch-over would always be greater than or equal to $\frac{T_S}{T_{G,\max}}$.


\section{Simulation}
\label{section:simulation}
We evaluate the proposed policy in VISSIM \cite{VISSIM}, which is a standard microscopic traffic simulator for transportation systems. 
In addition to the built-in features for conventional traffic signal control, VISSIM also provides programming integration with MATLAB to support user-customizable traffic control algorithms. 

\begin{figure}[H]
\begin{center}
\includegraphics[scale=0.17]{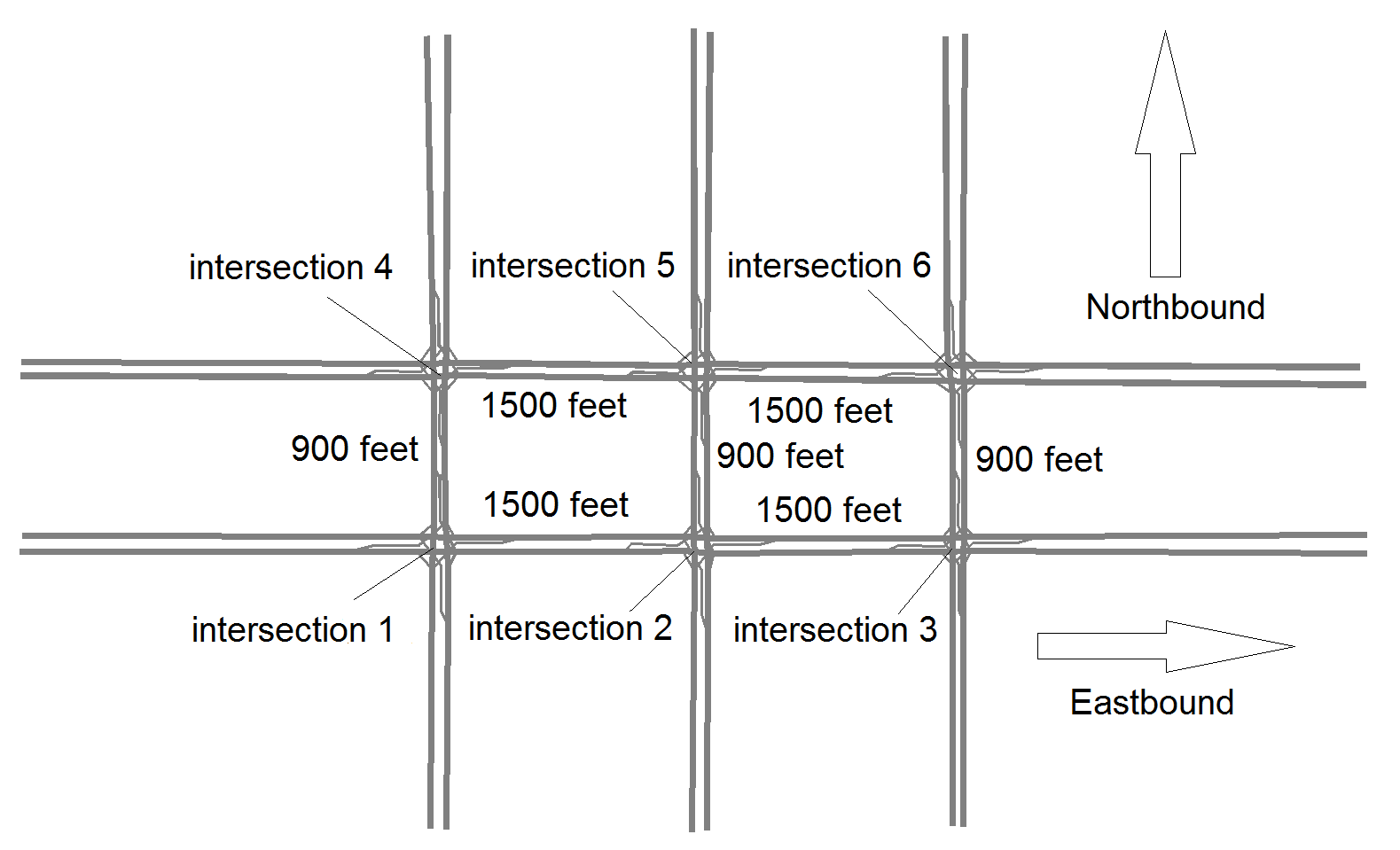}
\label{figure:VISSIM network}
\end{center}
\caption{System topology in VISSIM.}
\end{figure}

We consider a system of six signalized intersections as shown in Figure \ref{figure:VISSIM network}. 
In total, there are 10 entry links (4 major entries from the East and the West along with 6 minor entries from the North and the South) and 10 exit links.
Besides, the number of lanes of each through-traffic link and left-turn link are 3 and 1, respectively. 

According to the official statistics \cite{HCM2000}, the saturation flow of each link is set to be 1900 vehicles per hour per lane. 
Vehicles enter the system from the entry links and are routed towards an exit link in a probabilistic manner. 
We set the routing probability to be 0.2 and 0.8 for left-turn movement and through movement, respectively. 
We use $\lambda_{\text{E}}$, $\lambda_{\text{W}}$, $\lambda_{\text{N}}$, and $\lambda_{\text{S}}$ to denote the arrival rates of the entry links coming from the East, the West, the North, and the South, respectively.
We use the default driver behavior and lane-change model provided in VISSIM.
The speed limit of each vehicle is 40 miles per hour.
Each intersection has four admissible phases as described in Figure \ref{figure:intersection}. 
Throughout the simulation, we choose the slot time to be 1 second which is sufficient for updating the scheduling decisions.
The switch-over delay is set to be 5 seconds, which includes an amber period of 3 seconds and an all-red period of 2 seconds.
An important feature of our VISSIM simulation is that we consider the effect of finite buffer size.
When a link is fully occupied by vehicles, VISSIM will prohibit new vehicles, which can be either from the external or from upstream links, from joining the link and hence lower the throughput.

We compare the B-MP policy against the conventional fixed-time policy, Max-Pressure (MP) policy, and the Variable Frame-Based Max-Weight (VFMW) policy.
For the fixed-time policy, the timing plan is calculated by Synchro \cite{Synchro}, which is a widely-used optimization tool for timing plan design in transportation research. 
Throughout the simulation, we assume that the fixed-time policy has perfect information about the average traffic statistics of each link and therefore is able to optimize the timing plan accordingly. 
For VFMW, we choose the frame size to be $T_S+\bigl(\sum_{(i,j)\in\caM_{v}}Q_{i,j}(t_k)\bigr)^{0.9}$ as suggested in \cite{Celik2015}.  
For the B-MP policy, we choose $\alpha=0.01$ and $\beta=0.99$ as discussed in Section \ref{section:policy:proof}. Besides, to mitigate possible queue overflow due to finite queue capacity, we use weighted queue length with $q_{i,j}=3$ for through-traffic queues and $q_{i,j}=1$ for left-turn queues as discussed in Section \ref{section:variations:weighted}.
First, we consider the arrival traffic pattern as follows:
\vspace{1mm}

{\noindent \bf Scenario 1:}
$\lambda_{\text{E}}=\lambda_{\text{W}}=\bar{\lambda}$ and $\lambda_{\text{N}}=\lambda_{\text{S}}=0.5\cdot\bar{\lambda}$ (veh/hr). 

\begin{figure}[H]
\begin{center}
\includegraphics[scale=0.42]{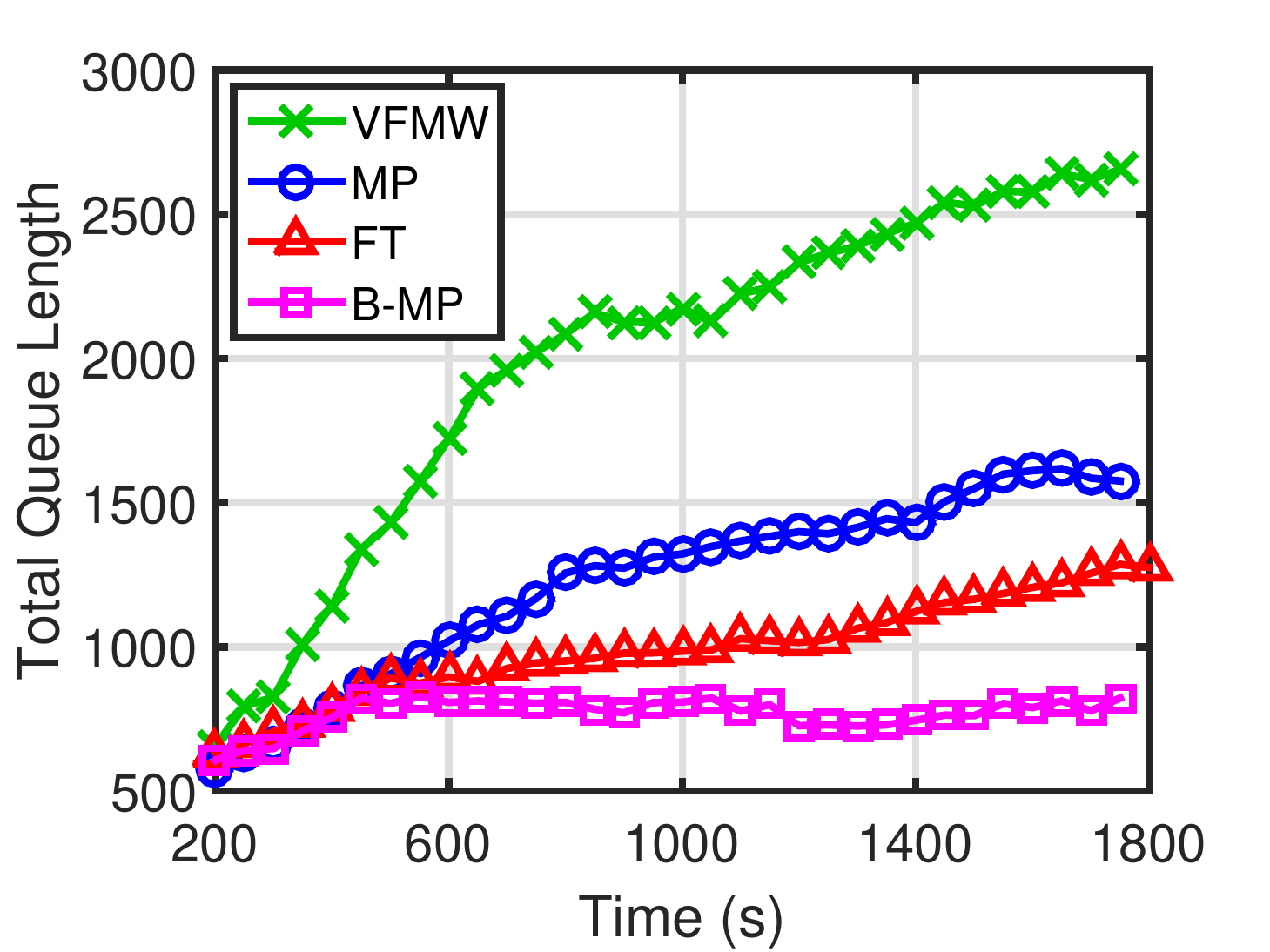}
\caption{Total queue length of the system under the four policies with $\bar{\lambda}=2400$.}
\label{figure:timeplot 2400}
\end{center}
\end{figure}

\vspace{1mm}
Under this traffic pattern, the maximum achievable $\bar{\lambda}$ is about 2600 veh/hr according to the traffic equations given by (\ref{equation:traffic equation matrix form}).
The total simulation time is 1800 seconds. 
Figure \ref{figure:timeplot 2400} shows the total number of vehicles in the system with $\bar{\lambda}=2400$ under the four policies.
We observe that B-MP indeed achieves the smallest total queue length while the total queue length keeps increasing under any of the other three policies.
\vspace{-3mm}
\vspace{-5mm}
\begin{figure}[H]
\begin{center}
\subfigure[System throughput]{
\includegraphics[scale=0.41]{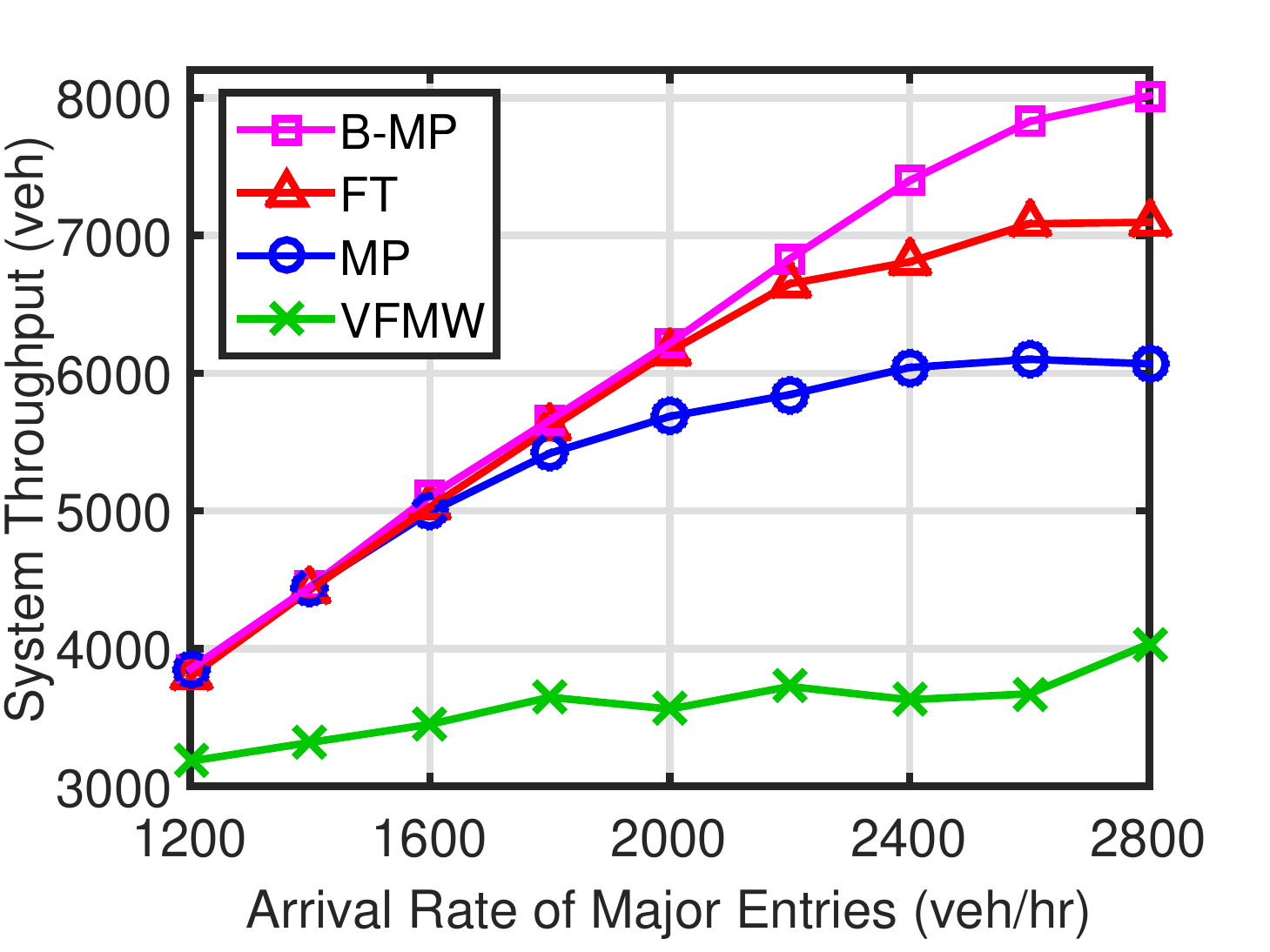}
\label{figure:thru}}
\subfigure[Average delay]{
\includegraphics[scale=0.41]{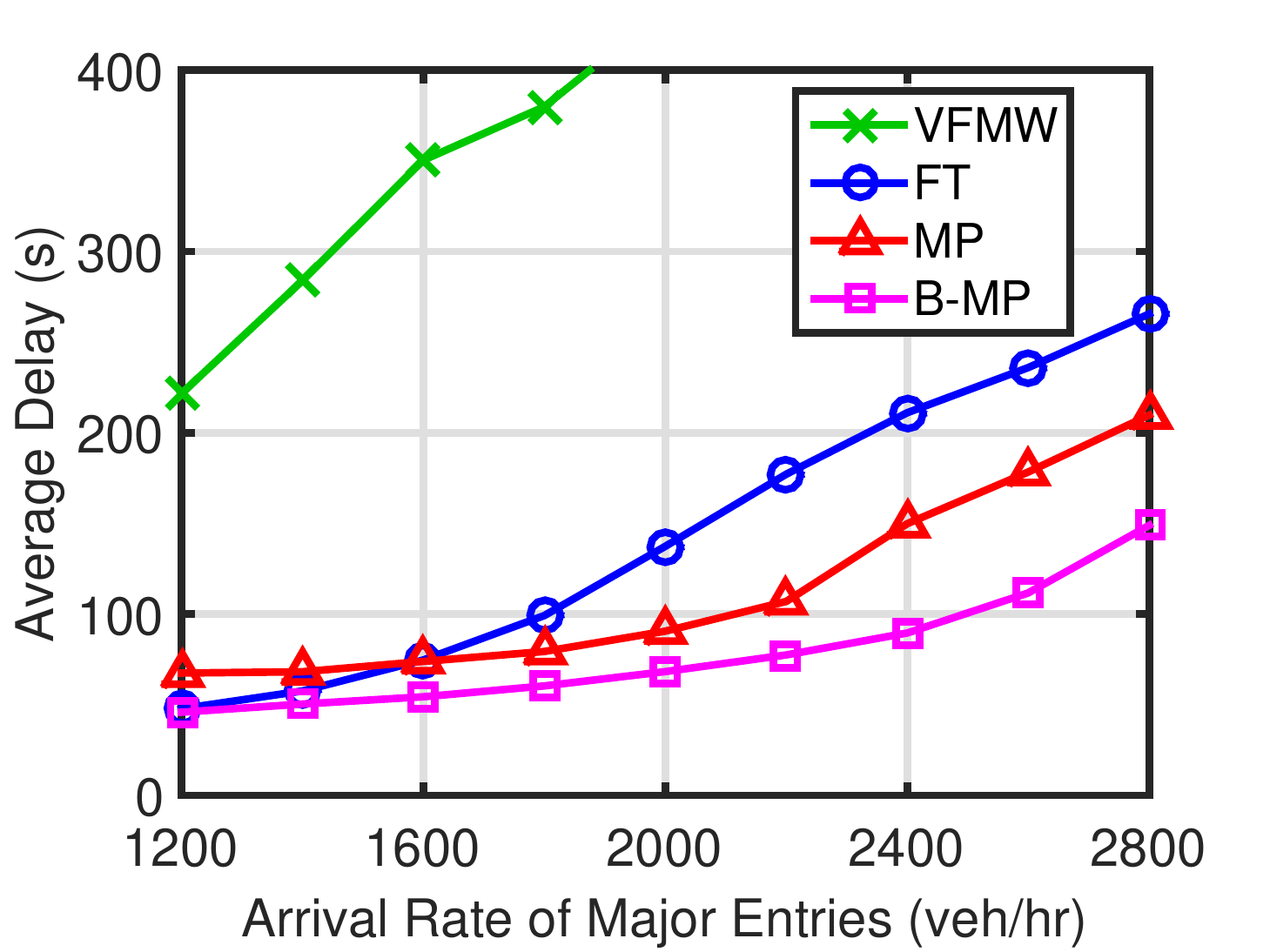}
\label{figure:delay}}
\end{center}
\caption{Delay and throughput performance under the four policies with different arrival rates.}
\end{figure}
Next, we measure the performance with $\bar{\lambda}$ between 1200 and 2800. 
Figure \ref{figure:thru} and Figure \ref{figure:delay} show the system throughput and average delay with different arrival rates.
Note that the average delay here is defined as the difference between the actual traveling time and the traveling time without any stop at the intersections.
In Figure \ref{figure:thru}, we see that under B-MP the throughput grows linearly with the arrival rate for $\bar{\lambda}$ up to 2600. With $\bar{\lambda}=2800$, the throughput under B-MP gets saturated simply because the arrival rate is already beyond the capacity region.
For the fixed-time policy, it can support $\bar{\lambda}$ only up to 2200 due to the capacity loss resulting from the switch-over delay.  
For MP and VFMW, they both suffer from severe capacity loss due to frequent switching of traffic signals. 
In Figure \ref{figure:delay}, the B-MP still achieves the smallest delay for every $\bar{\lambda}$.
For the heavy traffic condition with $\bar{\lambda}=2600$, compared to the fixed-time policy with perfect knowledge of traffic statistics, B-MP reduces the average delay by more than 40\% without any arrival rate information.  
For VFMW, we only show the average delay for $\bar{\lambda}$ below 1800 simple because it performs much more poorly than the other three policies for $\bar{\lambda}$ above 2000.


Next, we further consider time-varying arrival rates: 
\vspace{1mm}

{\noindent \bf Scenario 2:}  
\vspace{-2mm}
\begin{itemize}
\itemsep0em
\item \SI{0}{\second} to \SI{1200}{\second}: $(\lambda_{\text{W}}, \lambda_{\text{E}},\lambda_{\text{N}},\lambda_{\text{S}})=(2000,2000,1000,1000)$. 
\item \SI{1201}{\second} to \SI{2400}{\second}: $(\lambda_{\text{W}}, \lambda_{\text{E}},\lambda_{\text{N}},\lambda_{\text{S}})=(2500,1500,1500,500)$. 
\item \SI{2401}{\second} to \SI{3600}{\second}: $(\lambda_{\text{W}}, \lambda_{\text{E}},\lambda_{\text{N}},\lambda_{\text{S}})=(1500,2500,500,1500)$. 
\end{itemize}

\begin{figure}[H]
\begin{center}
\includegraphics[scale=0.42]{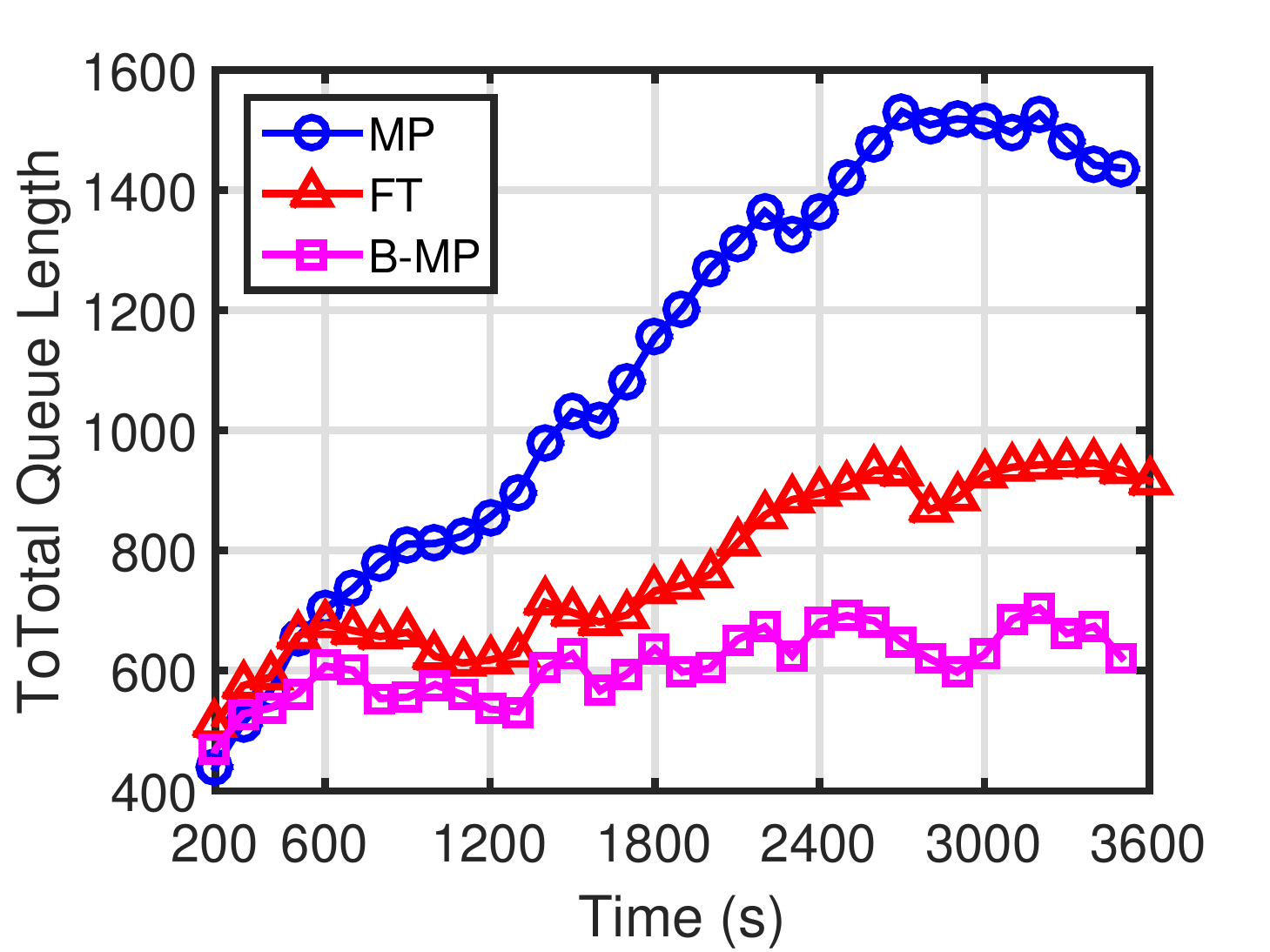}
\end{center}
\caption{Total queue length under time-varying traffic condition.}
\label{figure:time-varying}
\end{figure}

Note that the total arrival rate of the whole system remains the same under the above traffic pattern.
Figure \ref{figure:time-varying} shows that total queue length under the three policies. 
Here we omit the VFMW policy simply because it has much larger total queue length.
Again, B-MP still achieves the smallest total queue length at any time.
It is notable that the total queue length under B-MP does not change much with the time-varying pattern.
In contrast, the fixed-time policy suffers from much more congestion during time \SI{1200}{\second} to \SI{3600}{\second}.
This is because the fixed-time policy optimizes its timing plan based on the average arrival rates and thus fails to accommodate traffic dynamics.
Similar to Figure \ref{figure:timeplot 2400}, MP still performs quite poorly due to the service loss incurred by the switch-over delay.

\begin{figure}[H]
\begin{center}
\subfigure[System throughput]{
\includegraphics[scale=0.41]{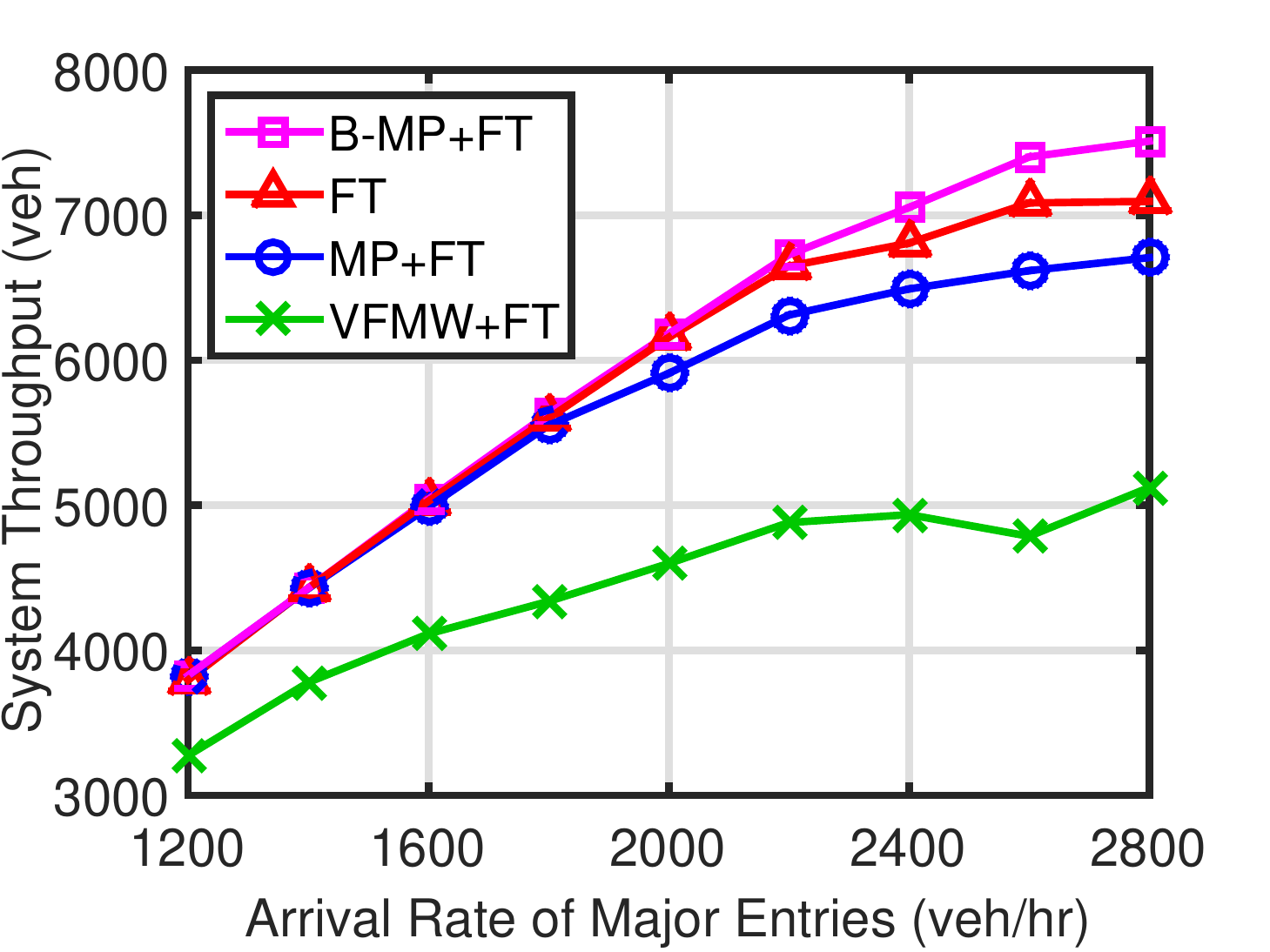}
\label{figure:thru mixed}}
\hspace{2mm}
\subfigure[Average delay]{
\includegraphics[scale=0.41]{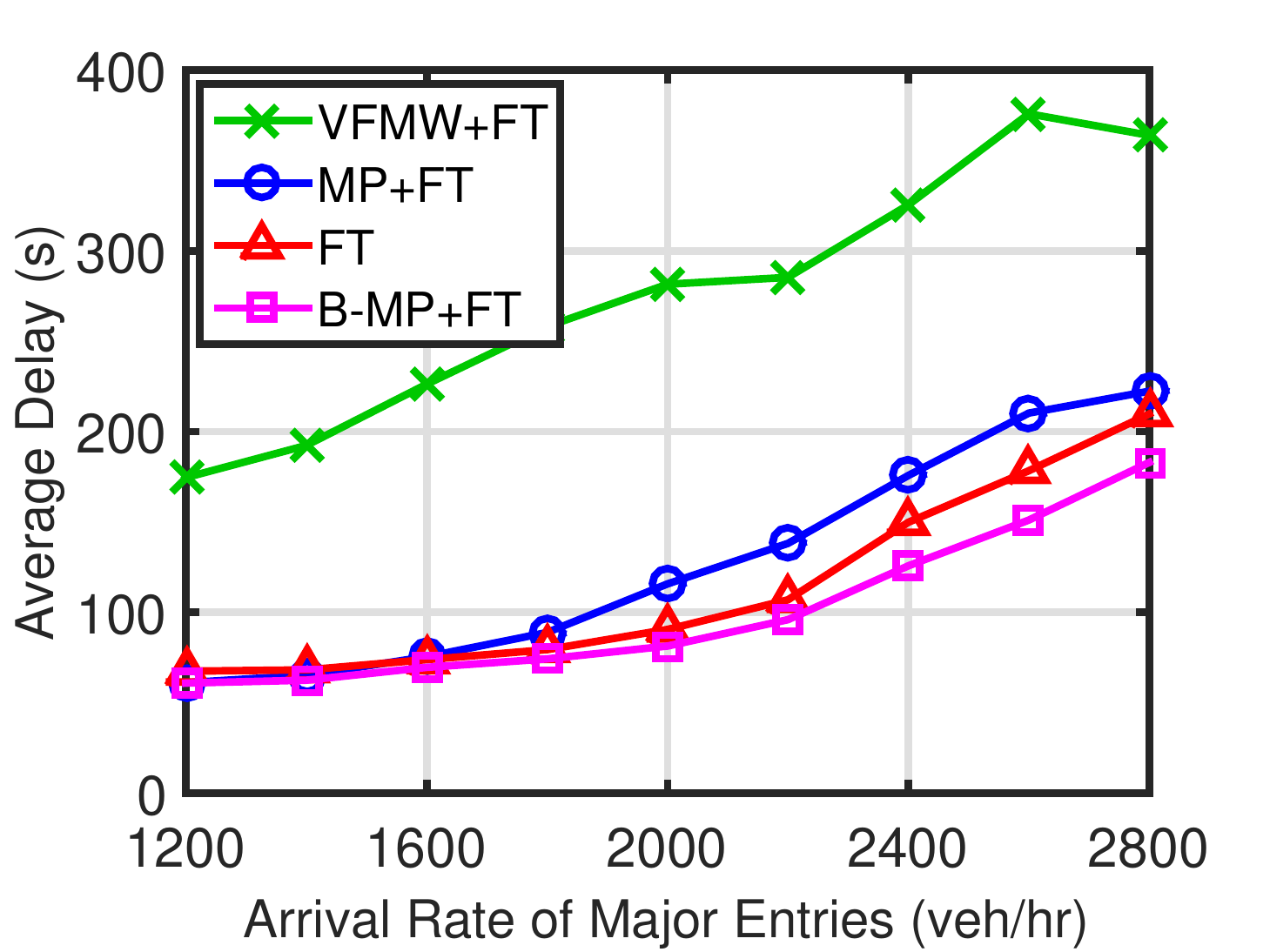}
\label{figure:delay mixed}}
\caption{Delay and throughput performance under the four policies in the partially-connected system.}
\end{center}
\end{figure}

Lastly, we consider a partially-connected system where three of the intersections are connected under a user-customized policy (B-MP, MP, or VFMW) and the rest are fixed-time intersections as usual.
Figure \ref{figure:thru mixed} and Figure \ref{figure:delay mixed} show the average delay and system throughput of the partially-connected system with different arrival rates.
Compared to the pure fixed-time system, partial inclusion of the B-MP policy still provides improvement in both throughput and average delay.
Besides, B-MP still outperforms the other two policies by a large margin in the partially-connected system. 
Through the above simulation, we demonstrate that B-MP indeed provides significant improvement over the other three popular policies.

\vspace{-2mm}
\section{Conclusion}	
\label{section:conclusion}
In this paper, we study the scheduling problem for networked transportation systems with switch-over delay. We propose a distributed scheduling policy that is throughput-optimal with switch-over delay without the knowledge of traffic demands. Moreover, the proposed policy still remains optimal when there are both fixed-time intersections and connected intersections. Hence, the proposed policy can still perform well in partially-connected systems. Simulation results show that the proposed policy indeed outperforms the other existing policies.

\vspace{-3mm}
\vspace{-6mm}

\vspace{-2mm}
\bibliographystyle{acm}
\bibliography{reference}
\appendix
\section{Proof of Lemma 1}
\label{appendix: proof of lemma 1}
\begin{proof}
Given $\mathrm{\Delta} L(t_k)=2{\bs{Q}}(t_k)^{\T}{\mathrm{\Delta}\bs{Q}(t_k)} + {\mathrm{ \Delta}\bs{Q}(t_k)}^{\T}{\mathrm{\Delta}\bs{Q}(t_k)}$, we provide an upper bound for each term separately. First, we derive an upper bound of $\E\sbkt[\big]{{\mathrm{\Delta}\bs{Q}(t_k)}^{\T}{\mathrm{\Delta}\bs{Q}(t_k)}\sgiven \bs{Q}(t_k)}$.
Define $V_{\max}:=\max\{A_{\max},S_{\max}\}$.
If a link $i$ is an entry link and $j\in \caD(i)$, then
\begin{equation}
\norm{\mathrm{\Delta} Q_{i,j}(t_k)} \leq T_k \max\{A_{\max},S_{\max}\}=T_kV_{\max}.
\end{equation} 
Otherwise, if $i\in\caL_{\text{int}}$ and $j\in \caD(i)$, then we know 
\begin{equation}
\lvert \mathrm{\Delta} Q_{i,j}(t_k)\rvert \leq U_{\max}T_kS_{\max}.
\end{equation}
Hence, we have
\begin{equation}
\E\sbkt[\big]{{\mathrm{\Delta}\bs{Q}(t_k)}^{\T}{\mathrm{\Delta}\bs{Q}(t_k)}\sgiven \bs{Q}(t_k)}\leq \norm{\caM}U_{\max}^{2} V_{\max}^{2}T_k^{2}.\label{equation: upper bound dQTdQ}
\end{equation}
Next, we provide an upper bound on $\E\sbkt[\big]{{\bs{Q}}(t_k)^{\T}{\mathrm{\Delta}\bs{Q}(t_k)}\sgiven \bs{Q}(t_k)}$.
\begin{align}
&{\bs{Q}}(t_k)^{\T}{\mathrm{\Delta}\bs{Q}(t_k)}\label{equation:drift QdQ start}\\
&=\sum_{t=t_k}^{t_{k+1}-1}\sum_{(i,j)\in \caM}\Biggl[-Q_{i,j}(t)\bkt[\Big]{S_{i,j}(t)I_{i,j}(t)X_{i,j}(t)\wedge Q_{i,j}(t)}\\
&\hspace{6pt}+\sum_{m:(m,i)}Q_{i,j}(t)\bkt[\Big]{S_{m,i}(t)I_{m,i}(t)X_{m,i}(t)\wedge Q_{i,j}(t)}R_{i,j}(t)\Biggr]\\
&\hspace{6pt}+\sum_{t=t_k}^{t_{k+1}-1}\sum_{i\in \caL_{\text{entry}}, j\in \caD(i)} Q_{i,j}(t_k)A_{i,j}(t)\\
&=\sum_{t=t_k}^{t_{k+1}-1}\sum_{(i,j)\in \caM}\Biggl[\bkt[\Big]{S_{i,j}(t)I_{i,j}(t)X_{i,j}(t)\wedge Q_{i,j}(t)}\times\\
&\hspace{36pt}\bkt[\Big]{-Q_{i,j}(t)+\sum_{p\in\caD(j)}R(j,p)(t)Q(j,p)(t_k)}\Biggr]\\
&\hspace{6pt}+\sum_{t=t_k}^{t_{k+1}-1}\sum_{i\in \caL_{\text{entry}}, j\in \caD(i)} Q_{i,j}(t_k)A_{i,j}(t).
\end{align}
Therefore, we have
\begin{align}
&\E\sbkt[\Big]{{\bs{Q}}(t_k)^{\T}{\mathrm{\Delta}\bs{Q}(t_k)}\sgiven\bs{Q}(t_k)}=\\
&-\sum_{t=t_k}^{t_{k+1}-1} \sum_{(i,j)\in \caM}\Biggl[\E\sbkt[\bigg]{ S_{i,j}(t)I_{i,j}(t)X_{i,j}(t)\wedge Q_{i,j}(t)\sgiven \bs{Q}(t_k)}\\
&\hspace{18pt}\times W_{i,j}(t_k)\Biggr]+ T_{k}\bkt[\bigg]{\sum_{i\in\caL_{\text{entry}},j\in\caD(i)}\lambda_{i}^{*}r_{i,j}Q_{i,j}(t_k)}.\label{equation:lambda rij Qij}
\end{align}
Moreover, we can rewrite the second term of (\ref{equation:lambda rij Qij}) as
\begin{align}
&\sum_{i\in\caL_{\text{entry}},j\in\caD(i)}\lambda_{i}^{*}r_{i,j}Q_{i,j}(t_k)\\
&=\sum_{i\in\caL_{\text{entry}},j\in\caD(i)}\lambda_{i}^{*}r_{i,j}Q_{i,j}(t_k)
+\sum_{j\in \caL_{\text{int}},p\in \caD(j)}\lambda_{j}^{*}r_{j,p}Q_{j,p}(t_k)\\
&\hspace{80pt}-\sum_{j\in \caL_{\text{int}},p\in \caD(j)}\lambda_{j}^{*}r_{j,p}Q_{j,p}(t_k)\\
&=\sum_{(i,j)\in\caM}\lambda_{i}^{*}r_{i,j}Q_{i,j}(t_k)-\sum_{j\in \caL_{\text{int}},p\in \caD(j)}\lambda_{j}^{*}r_{j,p}Q_{j,p}(t_k)\\
&=\sum_{(i,j)\in \caM}\lambda_{i}^{*}r_{i,j}\bkt[\Big]{Q_{i,j}(t_k)-\sum_{p:(j,p)}r_{j,p}Q_{j,p}(t_k)}\\
&=\sum_{(i,j)\in \caM}\lambda_{i}^{*}r_{i,j}W_{i,j}(t_k).
\end{align}
Therefore, we have
\begin{align}
&\E\sbkt[\Big]{{\bs{Q}}(t_k)^{\T}{\mathrm{\Delta}\bs{Q}(t_k)}\sgiven\bs{Q}(t_k)}\\
&=\sum_{t=t_k}^{t_{k+1}-1}\sum_{(i,j)\in\caM}\Biggl[W_{i,j}(t_k)\times\\
&\biggl(\lambda_{i}^{*}r_{i,j}-\E\sbkt[\Big]{S_{i,j}(t)I_{i,j}(t)X_{i,j}(t)\wedge Q_{i,j}(t)\sgiven\bs{Q}(t_k)}\biggr)\Biggr].\label{equation:expected QdQ}
\end{align}
We further decompose (\ref{equation:expected QdQ}) into two parts $\alpha_1$ and $\alpha_2$:
\begin{align}
&\alpha_1=\sum_{t=t_k}^{t_{k+1}-1}\sum_{(i,j)\in\caM}\Biggl[W_{i,j}(t_k)\times\label{equation:alpha 1 decompose 1}\\
&\hspace{60pt}\E\sbkt[\Big]{\lambda_{i}^{*}r_{i,j}-\mu_{i,j}I_{i,j}(t)X_{i,j}(t)\sgiven\bs{Q}(t_k)}\Biggr],\label{equation:alpha 1 decompose 2}\\
&\alpha_2=\sum_{t=t_k}^{t_{k+1}-1}\sum_{(i,j)\in\caM}\Biggl[W_{i,j}(t_k)\times\E \biggl[\mu_{i,j}I_{i,j}(t)X_{i,j}(t)\label{equation:alpha 2 decompose 1}\\
&\hspace{40pt}-\bkt[\big]{S_{i,j}(t)I_{i,j}(t)X_{i,j}(t)\wedge Q_{i,j}(t)}\biggr\vert\bs{Q}(t_k)\biggr]\Biggr].\label{equation:alpha 2 decompose 2}
\end{align}
We start with $\alpha_2$. For each movement $(i,j)$, if $Q_{i,j}(t)\geq S_{\max}$, then we know
\begin{equation}
\E\sbkt[\bigg]{\mu_{i,j}I_{i,j}(t)X_{i,j}(t)-\bkt[\big]{S_{i,j}(t)I_{i,j}(t)X_{i,j}(t)\wedge Q_{i,j}(t)}\sgiven\bs{Q}(t_k)}=0.\label{equation:alpha 2 simplify 1}
\end{equation}
Otherwise, if $Q_{i,j}(t)< S_{\max}$, we have
\begin{align}
&W_{i,j}(t_k)\cdot\E \biggl[\mu_{i,j}I_{i,j}(t)X_{i,j}(t)-\\
&\bkt[\big]{S_{i,j}(t)I_{i,j}(t)X_{i,j}(t)\wedge Q_{i,j}(t)\Bigr\vert\bs{Q}(t_k)}\biggr]\leq\mu_{i,j}S_{\max}\label{equation:alpha 2 simplify 2}
\end{align}
since we know $W_{i,j}(t_k)\leq Q_{i,j}(t_k)$ by definition. Therefore, we have the following upper bound of $\alpha_2$:
\begin{align}
\alpha_{2}&\leq \sum_{t=t_k}^{t_{k+1}-1}\sum_{(i,j)\in\caM}\mu_{i,j}S_{\max}\leq \bkt[\bigg]{\sum_{(i,j)\in \caM}\mu_{i,j}S_{\max}}T_k.\label{equation: upper bound of alpha2}
\end{align}
Note that (\ref{equation: upper bound of alpha2}) is an upper bound of $\alpha_2$ regardless of the scheduling policy.
Next, we consider $\alpha_1$.
Suppose $\bm{\lambda}$ is feasible, then by definition there must exist $\epsilon>0$ and $\bm{\Sigma}=(\Sigma_{i,j})$ in the convex hull of $\caI_v$ for each individual intersection $v\in \caV_{C}$ such that 
\begin{equation}
\mu_{i,j}\Sigma_{i,j}>\lambda_{i}^{*} r_{i,j}+\epsilon, \hspace{12pt} \forall (i,j)\in \caM_{v}.
\end{equation}
Moreover, we construct another vector $\bm{\Sigma}^{*}(t_k)=(\Sigma_{i,j}^{*}(t_k))$ as
\begin{align}
\Sigma_{i,j}^{*}(t_k)=\left \{ \begin{array}{rl}\frac{\lambda_{i}^{*} r_{i,j}+\epsilon}{\mu_{i,j}}, &\mbox{if $W_{i,j}(t_k)>0$}\\
0, &\mbox{otherwise}\end{array}\right.\label{equation:Sigma star}
\end{align}
Note that while $\bm{\Sigma}$ is a fixed vector across time,  we choose $\bm{\Sigma}^{*}(t_k)$ depending on the pressure $W_{i,j}(t_k)$.
It is easy to verify that $\bm{\Sigma}^{*}(t_k)$ is also in the convex hull of $\caI_v$.
For any max-pressure-at-switch-over policy, we must have that at time $t_k$
\begin{equation}
\sum_{(i,j)\in \caM_{v}}I_{i,j}(t_k)\mu_{i,j}W_{i,j}(t_k)\geq \sum_{(i,j)\in \caM_{v}}{\Sigma}^{*}_{i,j}(t_k)\mu_{i,j}W_{i,j}(t_k).
\end{equation}
Although the scheduling decision at $t_{k,l}^{v}$ with $l\geq 1$ cannot be determined purely by the information about $\bs{Q}(t_k)$, we still know that the pressure of the scheduled phase remains relatively large compared to the pressure of the scheduled phase at time $t_k$, i.e.
\begin{align}
&\sum_{(i,j)\in \caM_{v}}I_{i,j}(t_{k,l}^{v})\mu_{i,j}W_{i,j}(t_k)\label{equation:inequality of max-pressure start}\\
&\geq \sum_{(i,j)\in \caM_{v}}I_{i,j}(t_{k,l}^{v})\mu_{i,j}{W_{i,j}(t_{k,l}^{v})}\label{equation:inequality of max-pressure 0}\\
&\hspace{48pt}-\sum_{(i,j)\in \caM_{v}}(U_{\max}+1)V_{\max}I_{i,j}(t_{k,l}^{v})\mu_{i,j}(t_{k,l}^{v}-t_k)\label{equation:inequality of max-pressure 1}\\
&\geq \sum_{(i,j)\in \caM_{v}} \Sigma_{i,j}^{*}(t_k)\mu_{i,j}W_{i,j}(t_{k,l}^{v})\\
&\hspace{48pt}-\sum_{(i,j)\in \caM_{v}}(U_{\max}+1)V_{\max}I_{i,j}(t_{k,l}^{v})\mu_{i,j}(t_{k,l}^{v}-t_k)\label{equation:inequality of max-pressure 2}\\
&\geq \bkt[\bigg]{\sum_{(i,j)\in \caM_{v}}\Sigma_{i,j}^{*}(t_k)\mu_{i,j}W_{i,j}(t_k)}-C_0^{v}(t_{k,l}^{v}-t_k),\label{equation:inequality of max-pressure 3}
\end{align}
where $C_0^{v}=(U_{\max}+1)V_{\max}\bigl(\sum_{(i,j)\in \caM_{v}}\mu_{i,j}\bigr)$ is a positive constant.
(\ref{equation:inequality of max-pressure 0}) and (\ref{equation:inequality of max-pressure 1}) hold since $\norm{W_{i,j}(t+1)-W_{i,j}(t)}\leq (U_{\max}+1)V_{\max}$ for any $(i,j)$ and any $t$.
Note that $\alpha_1$ is a sum over all movements $(i,j)\in \caM$. Define 
\begin{align}
&F_{v}(t_k):=\sum_{t=t_k}^{t_{k+1}-1}\sum_{(i,j)\in\caM_{v}}\Biggl(W_{i,j}(t_k)\times\label{equation:Fv tk 0}\\
&\hspace{60pt}\E\biggl[\lambda_{i}^{*}r_{i,j}-\mu_{i,j}I_{i,j}(t)X_{i,j}(t)\Bigr\vert\bs{Q}(t_k)\biggr]\Biggr). 
\end{align}
Then, $\alpha_1=\sum_{v\in \caV}F_{v}(t_k)$.
For an intersection $v$ under any max-pressure-at-switch-over policy, by (\ref{equation:inequality of max-pressure start})-(\ref{equation:inequality of max-pressure 3}) we have
\begin{align}
F_{v}(t_k)&\leq T_S M^{v}_{k}\bkt[\Big]{\sum_{(i,j)\in\caM_{v}}\lambda_{i}^{*}r_{i,j}W_{i,j}(t_k)}\\
&+\bkt[\Big]{T_k-M_{k}^{v}T_S}C_0^{v}T_k\\
&+\bkt[\Big]{T_k-M_{k}^{v}T_S}\Biggl[-\epsilon \sum_{(i,j)\in \caM_{v}}\bkt[\big]{W_{i,j}(t_k)}^{+}\\
&\hspace{56pt}-\sum_{(i,j)\in \caM_{v}}\lambda_{i}^{*}r_{i,j}\bkt[\big]{W_{i,j}(t_k)}^{-}\Biggr].
\end{align}
Since $\lambda_{i}^{*}r_{i,j}<\mu_{i,j}\leq S_{\max}$, then for sufficiently small $\epsilon$ we have
\begin{align}
F_{v}(t_k)\leq& T_S M^{v}_{k}S_{\max}\bkt[\bigg]{{\sum_{(i,j)\in\caM_{v}}W_{i,j}(t_k)}}-\\
&\epsilon\bkt[\Big]{T_k-M_{k}^{v}T_S}\bkt[\bigg]{\sum_{(i,j)\in \caM_{v}}{W_{i,j}(t_k)}^{+}}+C_0^{v}T_k^{2}.\label{equation:upper bound alpha1 part 1}
\end{align}
On the other hand, for an intersection $v$ under the fixed-time control policy with cycle length $D_v$, we have
\begin{align}
&F_{v}(t_k)\\
&\leq \sum_{(i,j)\in M_{v}}{W_{i,j}(t_k)}^{+}\bkt[\Big]{T_k\lambda_{i}^{*}r_{i,j}-\mu_{i,j}\Sigma^{*}_{i,j}\xi_{v}(T_k-D_{v})}\\
&\hspace{12pt}-\sum_{(i,j)\in M_{v}}{W_{i,j}(t_k)}^{-}\bkt[\Big]{T_k\lambda_{i}^{*}r_{i,j}-\mu_{i,j}\Sigma^{*}_{i,j}\xi_{v}(T_k+D_{v})}\\
&=-\epsilon T_k \bkt[\bigg]{\sum_{(i,j)\in \caM_{v}}{W_{i,j}(t_k)}^{+}}- T_k \bkt[\bigg]{\sum_{(i,j)\in \caM_{v}}\lambda_{i}^{*}r_{i,j}{W_{i,j}(t_k)}^{-}}\\
&\hspace{12pt}+\sum_{(i,j)\in\caM_{v}}\mu_{i,j}\Sigma^{*}_{i,j}\xi_{v}D_{v}{W_{i,j}(t_k)}^{+}\\
&\leq -\epsilon T_k\bkt[\bigg]{\sum_{(i,j)\in \caM_{v}} {W_{i,j}(t_k)}^{+}}+\sum_{(i,j)\in\caM_{v}}\mu_{i,j}\Sigma^{*}_{i,j}\xi_{v}D_{v}{W_{i,j}(t_k)}^{+}\label{equation:upper bound alpha1 part 2}
\end{align}
where (\ref{equation:upper bound alpha1 part 2}) holds for sufficiently small $\epsilon$.
In summary, by the results in (\ref{equation: upper bound dQTdQ}), (\ref{equation: upper bound of alpha2}), (\ref{equation:upper bound alpha1 part 1}), (\ref{equation:upper bound alpha1 part 2}) as well as the fact that $W_{i,j}(t)^{+}\leq Q_{i,j}(t)$, we conclude that
\begin{align}
&\E\sbkt[\big]{\mathrm{\Delta} L(t_k)\sgiven \bs{Q}(t_k)}\leq -2\epsilon T_k \sum_{(i,j)\in\caM}{{W_{i,j}(t_k)}^{+}}\\
&+C_1 \sum_{v\in\caV_{C}}M_{k}^{v} \bkt[\bigg]{ \sum_{(i,j)\in\caM_{v}}{W_{i,j}(t_k)}^{+}}\\
&+C_2\sum_{v\in \caV_{{F}}}\sum_{(i,j)\in \caM_{v}}{W_{i,j}(t_k)}^{+}+C_3 T_k^{2}+C_4 T_k,
\end{align}
where 
\begin{align}
&C_1=2T_S \bkt[\big]{\epsilon+S_{\max}},\\
&C_2=2S_{\max}\bkt[\big]{\max_{v\in \caV_{F}}D_v},\\
&C_3=\lvert\caM \rvert U_{\max}^{2} V_{\max}^{2}+2\bkt[\Big]{\sum_{v\in\caV_{C}}C_0^{v}}\\
&C_4=2\bkt[\Big]{\sum_{(i,j)\in \caM}\mu_{i,j}S_{\max}}.
\end{align}
The proof is complete.$\hfill$
\end{proof}

\section{Proof of Lemma 2}
\label{appendix: proof of lemma 2}
\begin{proof}
Define $H(t_k):=\sum_{t=0}^{T_k-1}\sum_{(i,j)\in \caM}Q_{i,j}(t_k+t)$. Then, we have
\begin{align}
H(t_k)\leq& \sum_{t=0}^{T_k-1}\sum_{i\in \caL_{\textrm{entry}},j\in \caD(i)} \bkt[\Big]{Q_{i,j}(t_k)+\sum_{s=0}^{T_k-1}A_{i,j}(t_k+s)} \\
&+\sum_{t=0}^{T_k-1}\sum_{i\in \caL_{\textrm{int}},j\in \caD(i)} Q_{i,j}(t_k).
\end{align}
After taking conditional expectation of $H(t_k)$, we have
\begin{align}
&\E\sbkt[\big]{H(t_k)\sgiven \bs{Q}(t_k)}\\
&\leq T_k^{2} \bkt[\bigg]{\sum_{i\in\caL_{\text{entry}}, j\in \caD(i)} \lambda_i^{*}r_{i,j}} + T_k \bkt[\bigg]{\sum_{(i,j)\in \caM}Q_{i,j}(t_k)}\\
&\leq B_1 \bkt[\bigg]{\sum_{(i,j)\in\ \caM}Q_{i,j}(t_k)}^{1+\beta}
\end{align}
where $B_1=1+\sum_{i\in\caL_{\text{entry}}}\lambda_i^{*}r_{i,j}$ is a positive constant independent of $\bs{Q}(t_k)$. Then, by (\ref{equation:drift upper bound in lemma}),
\begin{align}
\E\sbkt[\big]{\mathrm{\Delta} L(t_{k})\sgiven \bs{Q}(t_k)}\leq B_0-\frac{\epsilon_0}{B_1}\E\sbkt[\big]{H(t_k)\sgiven \bs{Q}(t_k)}.\label{equation:drift upper bound by H}
\end{align}
By summing (\ref{equation:drift upper bound by H}) over all the superframes, we have
\begin{align}
\sum_{k\geq 0}\E\sbkt[\big]{\mathrm{\Delta} L(t_{k})\sgiven \bs{Q}(t_k)}\leq \sum_{k\geq 0} \bkt[\Big]{B_0-\frac{\epsilon_0}{B_1}\E\sbkt[\big]{H(t_k)\sgiven \bs{Q}(t_k)}}.\label{equation:drift upper bound by H all}
\end{align}
Given a finite initial condition $\bs{Q}(0)$, we have $L(0)<\infty$ and $\sum_{k\geq 0}\E\sbkt[\big]{\mathrm{\Delta} L(t_{k})\sgiven \bs{Q}(t_k)}\geq -L(0)$. Hence, we conclude that
\begin{align*}
\limsup_{T\rightarrow \infty}\frac{\sum_{t=0}^{T-1}\E\sbkt[\big]{\sum_{(i,j)\in\caM}Q_{i,j}(t)}}{T}\leq \frac{B_1\bkt[\big]{B_0+L(0)}}{\epsilon_0}< \infty.
\end{align*}
$\hfill$
\end{proof}

\section{Proof of Lemma 3}
\label{appendix: proof of lemma 3}
\begin{proof} 
We briefly summarize the results of Perron-Frobenius Theorem for non-negative matrices.
First, we consider the connection between a matrix and the induced directed graph.
\begin{definition}
Let $\bs{P}=(p_{ij})$ be an $n\times n$ matrix. The graph $G(\bs{P})$ is defined as the directed graph on $n$ nodes $\{N_1,...,N_{n}\}$ where there is a directed edge from $N_{i}$ to $N_{j}$ if and only if $p_{i,j}\neq 0$, for all $i,j=1,...,n$.
\end{definition}
\begin{definition}
Let $\bs{P}=(p_{ij})$ be an $n\times n$ matrix. $\bs{P}$ is said to be irreducible if and only if the corresponding directed graph $G(\bs{P})$ is strongly connected.
\end{definition}
\begin{lemma_md}(Perron-Frobenius Theorem \cite{Meyer2000})
\label{lemma:Perron-Frobenius}
Let $\bs{P}$ be an ${n\times n}$ irreducible non-negative matrix. There exists a unique strictly positive eigenvector $\bs{x}=(x_1,...,x_n)$ with $\sum_{i=1}^{n}{x_i}=1$ and the corresponding eigenvalue $\lambda_{\text{pf}}>0$.
\end{lemma_md}

Next, we consider the substochastic properties of a matrix.
\begin{definition}
A non-negative square matrix $\bs{P}$ is said to be substochastic if every row sum of $\bs{P}$ is less than or equal to 1 and at least one row sum is strictly less than 1. 
\end{definition}
\begin{lemma_md}(Section 9.4 in \cite{Hogben2006})
\label{lemma:substochastic eigenvalue less than 1}
Let $\bs{P}$ be an ${n\times n}$ substochastic matrix and $\bs{1}$ be an ${n\times n}$ identity matrix. Then, the maximum eigenvalue of $\bs{P}$ is less than 1 if and only if $(\bs{1}-\bs{P})$ is invertible.
\end{lemma_md}

Now, we are ready to prove Lemma \ref{lemma:sum W and sum Q inequality}.
Given a multi-hop system $\caG=(\caV, \caL, \caM)$, we construct another system $\caG'=(\caV{'}, \caL{'}, \caM{'})$ by adding both dummy links $\caL_{\text{dummy}}$ and dummy movements $\caM_{\text{dummy}}$ to the original system $\caG$. 
Specifically, for every node $v$ other than the source node $v_{\text{s}}$ and destination node $v_{\text{d}}$, we add an extra link from the $v_{\text{d}}$ to $v$.
For every dummy link $i$ from $v_{\text{d}}$ to $v$, we further add a new movement from link $i$ to link $j$ for every $j\in \caL$ with an associated dummy queue $Q_{i,j}$.
Therefore, for every dummy link $i$ from $v_{\text{d}}$ to $v$, we have $\caD(i)=\caL$.
Moreover, for every dummy link $i\in \caL_{\text{dummy}}$, the routing probability $r_{i,j}$ from link $i$ to a downstream link $j$ is set to be $\frac{1}{2|\caL|}$, for every $j\in \caD(i)$.
Therefore, $\sum_{j: j\in \caD(i),i\in \caL_{\text{dummy}}}=\frac{1}{2}<1$.
Therefore, the constructed graph $\caG'$ is strongly connected.
Let $\widetilde{\bs{Q}}=(Q_{i,j})_{(i,j)\in \caM'}$ and $\widetilde{\bs{W}}=(W_{i,j})_{(i,j)\in \caM'}$ be the corresponding queue length vector and pressure vector of the constructed system $\caG'$, respectively.
Then, we have 
\begin{equation}
\widetilde{\bs{W}} = (\bs{1}-\widetilde{\bs{R}})\widetilde{\bs{Q}},\label{equation:W tilde and Q tilde}
\end{equation}
where $\bs{1}$ is an $\lvert\caL' \rvert\times \lvert\caL' \rvert$ identity matrix and $\widetilde{\bs{R}}$ is the corresponding routing matrix with entries $\{r_{i,j}\}$ for all movements $(i,j)\in \caM'$.
It is easy to verify that $\widetilde{\bs{R}}$ is a substochastic matrix and $(\bs{1}-\widetilde{\bs{R}})$ is invertible by using elementary linear algebra. Note that for any dummy movement $(i,j)\in \caM'\setminus \caM$, we are allowed to freely assign values to $Q_{i,j}$ in any time slot and here we assign $Q_{i,j}=0$. Therefore, the corresponding $W_{i,j}$ is always non-positive, for any $(i,j)\in \caM'\setminus \caM$.

Since $\caG'$ is strongly connected, then the routing matrix $\widetilde{\bs{R}}$ is irreducible. 
By Lemma \ref{lemma:Perron-Frobenius}, there exists a unique strictly positive eigenvector $\bs{x}=(x_{i,j})_{(i,j)\in\caM'}$ of $\widetilde{\bs{R}}$ with $\sum_{(i,j)\in \caM'}x_{i,j} = 1$ and the corresponding eigenvalue $\lambda_{\text{pf}}>0$.
Hence, 
\begin{equation}
\bs{x}^{\T}\widetilde{\bs{R}}=\lambda_{\text{pf}}\bs{x}^{\T}.\label{equation:eigenvector of R tilde}
\end{equation}
Moreover, since $\widetilde{\bs{R}}$ is substochastic and $(\bs{1}-\widetilde{\bs{R}})$ is invertible, by Lemma \ref{lemma:substochastic eigenvalue less than 1} we also have $\lambda_{pf}<1$.
From (\ref{equation:W tilde and Q tilde}) and (\ref{equation:eigenvector of R tilde}), we have
\begin{align}
\bs{x}^{\T}\widetilde{\bs{W}}=\bs{x}^{\T}(\bs{1}-\widetilde{\bs{R}})\widetilde{\bs{Q}}=(1-\lambda_{\text{pf}})\bs{x}^{\T}\widetilde{\bs{Q}}
\end{align}
Therefore, $\bs{x}^{\T}\bkt[\Big]{\widetilde{\bs{W}}-(1-\lambda_{\text{pf}})\widetilde{\bs{Q}}}=0$. 
In other words, $\widetilde{\bs{W}}$ is in the perpendicular complement of the vector space spanned by $\bs{x}$. Hence, we can write $\widetilde{\bs{W}}$ as 
\begin{align}
&\widetilde{\bs{W}}=(1-\lambda_{\text{pf}})\widetilde{\bs{Q}}+\bs{y},
\end{align}
where $\bs{y}=(y_{i,j})_{(i,j)\in\caM'}$ is a vector orthogonal to the vector $\bs{x}$. Next, we consider two cases of $\bs{y}$:
\vspace{2mm}

\noindent {\bf Case 1:} $\bs{y}=\bs{0}$

Then, we directly have $\widetilde{\bs{W}}=(1-\lambda_{\text{pf}})\widetilde{\bs{Q}}$. Since $\widetilde{\bs{Q}}$ is non-negative, then $\widetilde{\bs{W}}$ is also non-negative. Moreover, since we assign $Q_{i,j}=0$ for every $(i,j)\in \caM'\setminus \caM$, then $W_{i,j}=0$, $\forall(i,j)\in \caM'\setminus \caM$. Hence, it is easy to verify that
\begin{align}
\sum_{(i,j)\in \caM}{W_{i,j}}^{+}=(1-\lambda_{\text{pf}})\sum_{(i,j)\in \caM}{Q_{i,j}}.\label{equation:sum W inequality 1}
\end{align}
\vspace{2mm}

\noindent {\bf Case 2:} $\bs{y}\neq\bs{0}$

Since $\bs{x}$ is strictly positive and $\bs{x}^{\T}\bs{y}=0$, then $\bs{y}$ cannot be non-positive. Let $(i_q^*, j_q^*)$ be the movement with the largest queue length in $\widetilde{\bs{Q}}$ and $(i_y^{*}, j_y^{*})$ be the movement with largest entry in $\bs{y}$.
Fix a small $\delta_1>0$. 
\begin{itemize}
\item If $W_{i_q^*, j_q^*}\geq \delta_1 Q_{i_q^*,j_q^*}$, then we have
\begin{align}
\sum_{(i,j)\in \caM}{W_{i,j}}^{+}\geq \delta_1  Q_{i_{q}^*,j_{q}^*}\geq \frac{\delta_1}{\lvert \caM\rvert}\sum_{(i,j)\in\caM}Q_{i,j}.\label{equation:sum W inequality 2}
\end{align}
\item If $W_{i_q^*, j_q^*}< \delta_1 Q_{i_q^*,j_q^*}$, then we know
\begin{equation}
y_{i_q^*,j_q^*}=W_{i_q^*,j_q^*}-(1-\lambda_{\text{pf}})Q_{i_q^*,j_q^*}<-(1-\lambda_{\text{pf}}-\delta_1)Q_{i_q^*,j_q^*}.
\end{equation}
Since $\bs{x}^{\T}\bs{y}=\sum_{(i,j)}x_{i,j}y_{i,j}=0$, then
\begin{equation}
x_{i_y^{*},j_y^{*}}y_{i_y^{*},j_y^{*}}\geq -\frac{y_{i_q^*,j_q^*}}{\lvert \caM'\rvert}>\frac{1}{\lvert \caM'\rvert}(1-\lambda_{\text{pf}}-\delta_1)Q_{i_q^*,j_q^*}. 
\end{equation}
Therefore, it is easy to verify that
\begin{align}
\sum_{(i,j)\in \caM}{W_{i,j}}^{+}\geq y_{i_y^{*},j_y^{*}}\geq\frac{1-\lambda_{\text{pf}}-\delta_1}{x_{i_y^{*},j_y^{*}} \lvert \caM'\rvert\lvert \caM\rvert}\sum_{(i,j)\in\caM}Q_{i,j}.\label{equation:sum W inequality 3}
\end{align}
\end{itemize}
In summary, by (\ref{equation:sum W inequality 1})-(\ref{equation:sum W inequality 3}), there always exists a constant $\delta>0$ such that $\sum_{(i,j)\in \caM}{W_{i,j}}^{+}\geq \delta\bkt[\Big]{\sum_{(i,j)\in \caM}{Q_{i,j}}}$. $\hfill$
\end{proof}

\section{Proof of Lemma 4}
\label{appendix: proof of lemma 4}
\begin{proof}
Consider the switch-over condition (\ref{equation:B-MP RHS}) in the $l$-th frame of the $k$-th superframe. For any $t\in[t_{k,l}^{v},t_{k,l+1}^{v}]$, the right-hand side of (\ref{equation:B-MP RHS}) is upper bounded as
\begin{align}
&\bkt[\bigg]{\sum_{(i,j)\in\caM_v}\mu_{i,j}{I}^{*}_{i,j}(t)W_{i,j}(t)}^{+}\\
\leq &\bkt[\bigg]{\sum_{(i,j)\in\caM_v}\mu_{i,j}{I}^{*}_{i,j}(t)W_{i,j}(t_{k,l}^{v})}^{+}+\norm{\caM_{v}}(U_{\max}+1)V_{\max}T_{k,l}^{v},
\end{align}
where $T^{v}_{k,l}:=t_{k,l+1}^{v}-t_{k,l}^{v}$, for all $k$ and $l$.
Similarly, we have a lower bound for the left-hand side of (\ref{equation:B-MP LHS}):
\begin{align}
&\bkt[\Big]{1+B_{v}(t_{k,l}^{v})}\biggl(\sum_{(i,j)\in\caM_v}\mu_{i,j}I^{*}_{i,j}(t_{k,l}^{v})W_{i,j}(t)\biggr)^{+}\\
\geq &\bkt[\Big]{1+B_{v}(t_{k,l}^{v})}\biggl(\sum_{(i,j)\in\caM_v}\mu_{i,j}I^{*}_{i,j}(t_{k,l}^{v})W_{i,j}(t_{k,l}^{v})\biggr)^{+}\\
& -\bkt[\Big]{1+B_{v}(t_{k,l}^{v})}\norm{\caM_{v}}(U_{\max}+1)V_{\max}T_{k,l}^{v}.
\end{align}
Therefore, since $B_{v}(t_{k,l}^{v})\leq \zeta T_S$, we have
\begin{align}
&{(2+\zeta T_S)\norm{\caM_{v}}(U_{\max}+1)V_{\max}}T_{k,l}^{v}\\
&\geq B_{v}(t_{k,l}^{v})\biggl(\sum_{(i,j)\in\caM_v}\mu_{i,j}I^{*}_{i,j}(t_{k,l}^{v})W_{i,j}(t_{k,l}^{v})\biggr)^{+}\\
&\geq B_{v}(t_{k,l}^{v})\cdot \max_{(i,j)\in\caM_v}\mu_{i,j}W_{i,j}(t_{k,l}^{v})^{+}\\
&\geq B_{v}(t_{k,l}^{v})\frac{\mu_{\min}}{\norm{\caM_v}}\biggl(\sum_{(i,j)\in\caM_v}W_{i,j}(t_{k,l}^{v})^{+}\biggr),
\end{align}
where $\mu_{\min}:=\min_{(i,j)\in\caM}\mu_{i,j}>0$.
Hence, we conclude that
\begin{align}
T_{k,l}^{v}\geq C_{5}B_{v}(t_{k,l}^{v})\biggl(\sum_{(i,j)\in\caM_v}W_{i,j}(t_{k,l}^{v})^{+}\biggr),
\end{align}
where $C_5={\mu_{\max}}\bkt[\Big]{(2+\zeta T_S)\norm{\caM_{v}}^{2}(U_{\max}+1)V_{\max}}^{-1}$.$\hfill$
\end{proof}

\section{Proof of Lemma 5}
\label{appendix: proof of lemma 5}
\begin{proof}
Consider the following two cases:
\vspace{2mm}

{\noindent \bf Case 1:} $\sum_{(i,j)\in\caM_v}W_{i,j}(t)^{+}<1$ for some $t\in[t_k, t_{k+1})$
\vspace{1mm}

Since $M_{k}^{v}\leq T_k$ and $\norm{W_{i,j}(t+1)-W_{i,j}(t)}\leq (U_{\max}+1)V_{\max}$ for any $(i,j)$ and any $t$, then we have
\begin{align*}
M_{k}^{v}\bkt[\Big]{\sum_{(i,j)\in \caM_{v}}{W_{i,j}(t_k)}^{+}}&<T_k\norm{\caM_{v}}\bkt[\Big]{1+(U_{\max}+1)V_{\max}T_k}\\
&\leq {C_6}\bkt[\Big]{\sum_{(i,j)\in \caM}Q_{i,j}(t_k)}^{2\beta},
\end{align*}
where $C_6=\norm{\caM_{v}}\bkt[\Big]{1+(U_{\max}+1)V_{\max}}$.
\vspace{2mm}

{\noindent \bf Case 2:} $\sum_{(i,j)\in\caM_v}W_{i,j}(t)^{+}\geq 1$ for all $t\in[t_k, t_{k+1})$
\vspace{1mm}

In this case, it is easy to verify that for any $t\in[t_k, t_{k+1})$,
\begin{align}
{\min\Biggl\{1, \biggl(\Bigl[\sum_{(i,j)\in\caM_v}W_{i,j}(t)\Bigr]^{+}\biggr)^{-\alpha}\Biggr\}}\geq \biggl(\sum_{(i,j)\in\caM_v}W_{i,j}(t)^{+}\biggr)^{-\alpha}.
\end{align}
Therefore, at each $t_{k,l}^{v}$ the bias function is lower bounded as
\begin{align}
B_{v}(t_{k,l}^{v})\geq \zeta T_{S}\biggl(\sum_{(i,j)\in\caM_v}W_{i,j}(t_{k,l}^{v})^{+}\biggr)^{-\alpha}.
\end{align}
By Lemma \ref{lemma:Tkl lower bound under B-MP}, we have
\begin{align}
\label{equation:Tkl lower bound in terms of Wij+}
T_{k,l}^{v}&\geq C_5 \zeta T_S\biggl(\sum_{(i,j)\in\caM_v}W_{i,j}(t_{k,l}^{v})^{+}\biggr)^{1-\alpha}\\
&\geq C_5 \zeta T_S\biggl[\sum_{(i,j)\in\caM_v}\Bigl(W_{i,j}(t_{k})^{+}-{(U_{\max}+1)V_{\max}T_k}\Bigr)^{+}\biggr]^{1-\alpha}\\
&\geq C_5 \zeta T_S \Biggl[\frac{1}{2^{1-\alpha}}\biggl(\sum_{(i,j)\in\caM_v}W_{i,j}(t_{k})^{+}\biggr)^{1-\alpha}\\
&{\hspace{90pt}-\bkt[\Big]{(U_{\max}+1)V_{\max}T_k}^{1-\alpha}\Biggr]}
\end{align}
where the last inequality holds since $\norm{a+b}^{p}\leq 2^p\bkt[\big]{\norm{a}^p+\norm{b}^p}$, for any $a,b\in \mathbb{R}$ and for any $p>0$.
Next, we need to discuss the following two possible scenarios:
\vspace{1mm}

{\noindent \bf Case 2-1:} 
\begin{equation}
\frac{1}{2^{1-\alpha}}\biggl(\sum_{(i,j)\in\caM_v}W_{i,j}(t_{k})^{+}\biggr)^{1-\alpha}\geq 2 \bkt[\Big]{(U_{\max}+1)V_{\max}T_k}^{1-\alpha}
\end{equation}

\vspace{1mm}

Then, we have a lower bound on $T_{k,l}^{v}$ as
\begin{align}
T_{k,l}^{v}\geq C_5 \zeta T_S\bkt[\Big]{(U_{\max}+1)V_{\max}T_k}^{1-\alpha}
\end{align}
Without loss of generality, we only need to consider the case where $C_5 \zeta T_S\bkt[\Big]{(U_{\max}+1)V_{\max}T_k}^{1-\alpha}>1$ (Otherwise, $T_k$ is upper bounded by a constant). 
Therefore, we have
\begin{align}
&M_{k}^{v}\bkt[\Big]{\sum_{(i,j)\in \caM_{v}}{W_{i,j}(t_k)}^{+}}\\
&\leq \frac{T_k}{C_5 \zeta T_S\bkt[\Big]{(U_{\max}+1)V_{\max}T_k}^{1-\alpha}-1}\cdot\bkt[\bigg]{\sum_{(i,j)\in \caM_{v}}{W_{i,j}(t_k)}^{+}}.
\end{align}
Since $\bkt[\Big]{\sum_{(i,j)\in \caM_{v}}{W_{i,j}(t_k)}^{+}}\leq \bkt[\Big]{\sum_{(i,j)\in \caM}{Q_{i,j}(t_k)}}$, there exists a constant $C_7>0$ such that
\begin{align}
M_{k}^{v}\bkt[\Big]{\sum_{(i,j)\in \caM_{v}}{W_{i,j}(t_k)}^{+}}&\leq  C_7 T_k^{\alpha}{\sum_{(i,j)\in \caM}{Q_{i,j}(t_k)}}\\
&=C_7 \biggl({\sum_{(i,j)\in \caM}{Q_{i,j}(t_k)}}\biggr)^{1+\alpha\beta}.
\end{align}

{\noindent \bf Case 2-2:} 
\begin{equation}
\frac{1}{2^{1-\alpha}}\biggl(\sum_{(i,j)\in\caM_v}W_{i,j}(t_{k})^{+}\biggr)^{1-\alpha}< 2 \bkt[\Big]{(U_{\max}+1)V_{\max}T_k}^{1-\alpha}
\end{equation}
\vspace{1mm}

In this case, it is easy to verify that there exists a constant $C_8>0$ such that
\begin{align}
&M_{k}^{v}\bkt[\Big]{\sum_{(i,j)\in \caM_{v}}{W_{i,j}(t_k)^{+}}}\leq C_8T_k^{2}=C_8\bkt[\Big]{\sum_{(i,j)\in \caM}Q_{i,j}(t_k)}^{2\beta}.
\end{align}
Therefore, the proof is complete.$\hfill$
\end{proof}

\section{Proof of Lemma 6}
\label{appendix: proof of Lemma 6}
\begin{proof}
We still consider the Lyapunov function $L({\bs{Q}}(t))=\sum_{(i,j)\in \caM}Q_{i,j}(t)^{2}$. 
By the condition that $Q_{i,j}(t)-Q_{i,j}^{\dagger}(t)\in[-B, B]$, we also have $W_{i,j}(t)-W_{i,j}^{\dagger}(t)\in [-2B,2B]$, for all $(i,j)$ and all $t$. 
Similar to the proof of Lemma \ref{lemma:drift upper bound}, we consider the conditional drift over one superframe:
\begin{equation}
\E\sbkt[\Big]{\mathrm{\Delta} L(t_k)\given{\bs{Q}}^{\dagger}(t_k)}=\E\sbkt[\Big]{2{\bs{Q}}(t_k)^{\T}{\mathrm{\Delta}\bs{Q}(t_k)} + {\mathrm{ \Delta}\bs{Q}}^{\T}{\mathrm{\Delta}\bs{Q}(t_k)}\given {\bs{Q}}^{\dagger}(t_k)}.
\end{equation}
First, similar to (\ref{equation: upper bound dQTdQ}), we have
\begin{align}
&\E\sbkt[\big]{\mathrm{\Delta}{\bs{Q}}(t_k)^{\T}{\mathrm{\Delta}\bs{Q}(t_k)}\given{\bs{Q}}^{\dagger}(t_k)}\leq \norm{\caM}U_{\max}^{2} V_{\max}^{2}T_k^{2}.
\end{align}
Next, we consider $\E\sbkt[\big]{{\bs{Q}}(t_k)^{\T}{\mathrm{\Delta}\bs{Q}(t_k)}\given{\bs{Q}}^{\dagger}(t_k)}$. Following the same procedure as in (\ref{equation:drift QdQ start})-(\ref{equation:expected QdQ}), we have
\begin{align}
&\E\Bigl[{{\bs{Q}}(t_k)^{\T}{\mathrm{\Delta}\bs{Q}(t_k)}\given{\bs{Q}}^{\dagger}(t_k)}\Bigr]\\
&=\sum_{t=t_k}^{t_{k+1}-1}\sum_{(i,j)\in\caM}\Biggl[\lambda_{i}^{*}r_{i,j}\E\Bigl[W_{i,j}(t_k)\given \bs{Q}^{\dagger}(t_k)\Bigr]\label{equation:expected QdQ dagger 1}\\
&\hspace{12pt}-\E\sbkt[\Big]{S_{i,j}(t)I_{i,j}(t)X_{i,j}(t)\wedge Q_{i,j}(t)\sgiven\bs{Q}^{\dagger}(t_k)}\Biggr]\label{equation:expected QdQ dagger 2}\\
&\leq \sum_{t=t_k}^{t_{k+1}-1}\sum_{(i,j)\in\caM}\Biggl[\lambda_{i}^{*}r_{i,j}\E\Bigl[W_{i,j}^{\dagger}(t_k)\given \bs{Q}^{\dagger}(t_k)\Bigr]\label{equation:expected QdQ dagger 3}\\
&\hspace{12pt}-\E\sbkt[\Big]{S_{i,j}(t)I_{i,j}(t)X_{i,j}(t)\wedge Q_{i,j}(t)\sgiven\bs{Q}^{\dagger}(t_k)}\Biggr]\label{equation:expected QdQ dagger 4}\\
&\hspace{12pt}+\norm{\caM}A_{\max}BT_k + \norm{\caM}S_{\max}BT_k.\label{equation:expected QdQ dagger 5}
\end{align}
Now, as in (\ref{equation:alpha 1 decompose 1})-(\ref{equation:alpha 2 decompose 2}), we further decompose (\ref{equation:expected QdQ dagger 3})-(\ref{equation:expected QdQ dagger 4}) into two parts:
\begin{align}
&\alpha^{\dagger}_1=\sum_{t=t_k}^{t_{k+1}-1}\sum_{(i,j)\in\caM}\Biggl[W_{i,j}^{\dagger}(t_k)\times\label{equation:alpha dagger 1 decompose 1 }\\
&\hspace{60pt}\E\sbkt[\Big]{\lambda_{i}^{*}r_{i,j}-\mu_{i,j}I_{i,j}(t)X_{i,j}(t)\sgiven\bs{Q}^{\dagger}(t_k)}\Biggr],\label{equation:alpha dagger 1 decompose 2}\\
&\alpha^{\dagger}_2=\sum_{t=t_k}^{t_{k+1}-1}\sum_{(i,j)\in\caM}\Biggl[W_{i,j}^{\dagger}(t_k)\times\E \biggl[\mu_{i,j}I_{i,j}(t)X_{i,j}(t)\label{equation:alpha dagger 2 decompose 1}\\
&\hspace{40pt}-\bkt[\big]{S_{i,j}(t)I_{i,j}(t)X_{i,j}(t)\wedge Q_{i,j}(t)}\biggr\vert\bs{Q}^{\dagger}(t_k)\biggr]\Biggr].\label{equation:alpha dagger 2 decompose 2}
\end{align}
By a similar argument as in (\ref{equation:alpha 2 simplify 1})-(\ref{equation:alpha 2 simplify 2}), we know
\begin{align}
\alpha^{\dagger}_{2}&\leq \bkt[\bigg]{\sum_{(i,j)\in \caM}\mu_{i,j}(S_{\max}+B)}T_k\label{equation: upper bound of alpha2}
\end{align}
since $W^{\dagger}_{i,j}(t_k)\leq Q^{\dagger}_{i,j}(t_k)\leq Q_{i,j}(t_k)+B$, for any $(i,j)$.
To calculate $\alpha^{\dagger}_1$, as in (\ref{equation:Sigma star}) we construct a vector $\bm{\Sigma}^{**}(t_k)=(\Sigma_{i,j}^{**}(t_k))$ as
\begin{align}
\Sigma_{i,j}^{**}(t_k)=\left \{ \begin{array}{rl}\frac{\lambda_{i}^{*} r_{i,j}+\epsilon}{\mu_{i,j}}, &\mbox{if $W_{i,j}^{\dagger}(t_k)>0$}\\
0, &\mbox{otherwise}\end{array}\right.
\end{align}
Again, by the max-pressure-at-switch-over property, we have
\begin{equation}
\sum_{(i,j)\in \caM_{v}}I_{i,j}(t_k)\mu_{i,j}W^{\dagger}_{i,j}(t_k)\geq \sum_{(i,j)\in \caM_{v}}{\Sigma}^{**}_{i,j}(t_k)\mu_{i,j}W^{\dagger}_{i,j}(t_k).
\end{equation}
Following the same procedure as in (\ref{equation:inequality of max-pressure start})-(\ref{equation:inequality of max-pressure 3}), we have
\begin{align}
&\sum_{(i,j)\in \caM_{v}}I_{i,j}(t_{k,l}^{v})\mu_{i,j}W^{\dagger}_{i,j}(t_k)\label{equation:inequality of max-pressure dagger start}\\
&\geq \bkt[\bigg]{\sum_{(i,j)\in \caM_{v}}\Sigma_{i,j}^{**}(t_k)\mu_{i,j}W^{\dagger}_{i,j}(t_k)}-C_0^{\dagger,v}(t_{k,l}^{v}-t_k),\label{equation:inequality of max-pressure dagger 1}
\end{align}
where $C_0^{\dagger,v}=(4B+(U_{\max}+1)V_{\max})\cdot\bigl(\sum_{(i,j)\in \caM_{v}}\mu_{i,j}\bigr)$.
Following the same discussion as in (\ref{equation:Fv tk 0})-(\ref{equation:upper bound alpha1 part 2}) with $W_{i,j}(t_k)$ replaced by $W^{\dagger}_{i,j}(t_k)$, we have
\begin{align}
&\E\sbkt[\big]{\mathrm{\Delta} L(t_k)\sgiven \bs{Q}^{\dagger}(t_k)}\leq -2\epsilon T_k \sum_{(i,j)\in\caM}{{W_{i,j}^{\dagger}(t_k)}^{+}}\\
&+C^{\dagger}_1 \sum_{v\in\caV_{C}}M_{k}^{v} \bkt[\bigg]{ \sum_{(i,j)\in\caM_{v}}{W_{i,j}^{\dagger}(t_k)}^{+}}\\
&+C^{\dagger}_2\sum_{v\in \caV_{{F}}}\sum_{(i,j)\in \caM_{v}}{W_{i,j}^{\dagger}(t_k)^{+}}+C^{\dagger}_3 T_k^{2}+C^{\dagger}_4 T_k,
\end{align}
where $C^{\dagger}_1,C^{\dagger}_2,C^{\dagger}_3,$ and $C^{\dagger}_4$ are some positive constants.$\hfill$
\end{proof}


\end{document}